\RequirePackage{rotating}
\documentclass[acmsmall]{acmart}
\usepackage{CJKutf8}
\usepackage{subfigure}
\usepackage{tablefootnote}
\usepackage{algorithm}
\usepackage{algorithmic}
\usepackage{multirow}
\usepackage{xcolor}

\AtBeginDocument{%
  \providecommand\BibTeX{{%
    \normalfont B\kern-0.5em{\scshape i\kern-0.25em b}\kern-0.8em\TeX}}}

\setcopyright{acmcopyright}
\copyrightyear{2023}
\acmYear{2023}
\acmDOI{10.1145/xxxxxxx.xxxxxxx}

\acmJournal{JACM}
\acmVolume{37}
\acmNumber{4}
\acmArticle{111}
\acmMonth{1}

\begin{document}

\title{Personalized Prompt Learning for Explainable Recommendation}


\author{Lei Li}
\affiliation{%
	\institution{Hong Kong Baptist University}
	\streetaddress{34 Renfrew Road}
	\city{Hong Kong}
	\country{China}
}
\email{csleili@comp.hkbu.edu.hk}

\author{Yongfeng Zhang}
\affiliation{%
	\institution{Rutgers University}
	\streetaddress{110 Frelinghuysen Road}
	\city{Piscataway}
	\state{New Jersey}
	\country{USA}
	\postcode{08854-8019}
}
\email{yongfeng.zhang@rutgers.edu}

\author{Li Chen}
\affiliation{%
	\institution{Hong Kong Baptist University}
	\streetaddress{34 Renfrew Road}
	\city{Hong Kong}
	\country{China}
}
\email{lichen@comp.hkbu.edu.hk}

\renewcommand{\shortauthors}{Li, Zhang, and Chen}

\begin{abstract}
Providing user-understandable explanations to justify recommendations could help users better understand the recommended items, increase the system's ease of use, and gain users' trust. A typical approach to realize it is natural language generation. However, previous works mostly adopt recurrent neural networks to meet the ends, leaving the potentially more effective pre-trained Transformer models under-explored. In fact, user and item IDs, as important identifiers in recommender systems, are inherently in different semantic space as words that pre-trained models were already trained on. Thus, how to effectively fuse IDs into such models becomes a critical issue. Inspired by recent advancement in prompt learning, we come up with two solutions: find alternative words to represent IDs (called discrete prompt learning), and directly input ID vectors to a pre-trained model (termed continuous prompt learning). In the latter case, ID vectors are randomly initialized but the model is trained in advance on large corpora, so they are actually in different learning stages. To bridge the gap, we further propose two training strategies: sequential tuning and recommendation as regularization. Extensive experiments show that our continuous prompt learning approach equipped with the training strategies consistently outperforms strong baselines on three datasets of explainable recommendation.
\end{abstract}

\begin{CCSXML}
	<ccs2012>
	<concept>
	<concept_id>10002951.10003317.10003347.10003350</concept_id>
	<concept_desc>Information systems~Recommender systems</concept_desc>
	<concept_significance>500</concept_significance>
	</concept>
	<concept>
	<concept_id>10010147.10010178.10010179.10010182</concept_id>
	<concept_desc>Computing methodologies~Natural language generation</concept_desc>
	<concept_significance>500</concept_significance>
	</ccs2012>
\end{CCSXML}

\ccsdesc[500]{Information systems~Recommender systems}
\ccsdesc[500]{Computing methodologies~Natural language generation}

\keywords{Explainable Recommendation; Transformer; Pre-trained Language Model; Prompt Learning}

\maketitle

\section{Introduction} \label{sec:intro}

\begin{figure}
	\centering
	\includegraphics[scale=0.5]{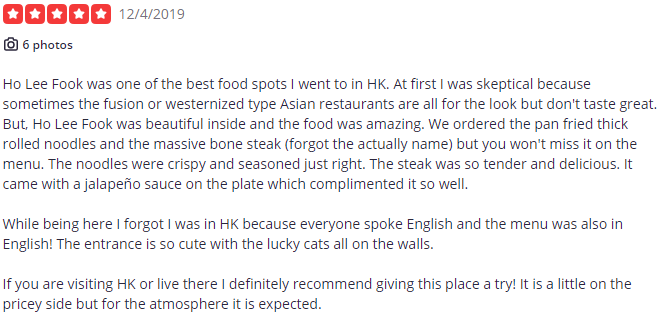}
	\caption{A review example from Yelp. The user and the restaurant are omitted for privacy protection.}
	\label{fig:example}
\end{figure}

Traditional recommender systems help users overcome the information overload problem by providing personalized recommendations (e.g., movies or songs) that cater to their interests.
Meanwhile, explanations that justify why these recommendations are made are becoming more and more important, as they can help users make better and faster decisions, increase the system's ease of use, and gain their trust in the system \cite{Handbook15-Explanation, FTIR20-Survey}.
There is a variety of explanation style, such as pre-defined templates \cite{SIGIR14-EFM, CIKM21-Counterfactual, JIIS21-CAESAR}, highlighted image regions \cite{SIGIR19-VECF} and automatically generated sentences \cite{EARS19-HSS, CIKM20-NETE, ACL21-PETER}.
The last type has gained increasing attention recently, mainly due to the availability of textual data on online commercial platforms, such as Amazon and Yelp, which encourage users to express their opinions by writing reviews (see Fig. \ref{fig:example}), as well as the advancement of natural language generation techniques, such as Recurrent Neural Networks (RNN), Transformer \cite{NIPS17-Transformer} and pre-trained language models \cite{18-GPT, NAACL19-BERT, NeurIPS19-UNILM}.

In particular, recent years have witnessed the stronger and stronger language modeling capability of large pre-trained models.
Taking Generative Pre-Training (GPT) series \cite{18-GPT, 19-GPT2, NeurIPS20-GPT3} as an example, the first generation GPT \cite{18-GPT} after fine-tuning achieves the state-of-the-art in 9 natural language understanding tasks out of 12.
Further, GPT-2 \cite{19-GPT2} without fine-tuning is able generate news articles that resemble authentic ones.
More surprisingly, GPT-3 \cite{NeurIPS20-GPT3} could even do simple arithmetic (e.g., 2 digit multiplication) that the model was not trained or fine-tuned for.
In the meantime, the size of these models and the volume of training data are becoming prohibitively large.
Regarding model size, GPT has 117 million parameters, while GPT-2 and GPT-3 are increased dramatically to 1.5 billion and 175 billion, respectively.
With respect to data, GPT takes as input 7000 books (approximately 7GB if a book has the size of 1MB), while GPT-2 and GPT-3 are fed 40GB and 570GB textual data, respectively.

As a consequence, it is nearly impossible to do customized modifications on the structure of these models.
Moreover, it would also be challenging to incorporate into them user and item IDs, which are indispensable in recommender systems but are in very different semantic space as words that these models were trained on.
No wonder most previous works \cite{EARS19-HSS, IJCAI19-CAML, CIKM20-NETE, WSDM21-SAER, sun2021unsupervised} adopt RNN, such as Long Short-Term Memory (LSTM) \cite{Neural97-LSTM} and Gated Recurrent Unit (GRU) \cite{EMNLP14-GRU}, or small unpretrained Transformer \cite{ACL21-PETER} for explanation generation.
This, however, makes the more effective pre-trained models less explored.

Fortunately, recent progress made in prompt learning \cite{CSUR22-Survey} points out a promising way.
Instead of modifying the structure of pre-trained models, researchers seek to adapt a given task to the models, so that they can directly model text probability.
For instance, a prompt for sentiment classification could be constructed with the format of ``\underline{I love this book.} This book is'', where the underlined text is a specific sample and the remaining words are a hand-crafted template.
This type of conditioning textual string is referred to as \textit{discrete prompt}.
After feeding it to a pre-trained model, a word prediction can be made at the end of the string, such as ``good'' or ``bad'', indicating a positive or negative sentiment.

Likewise, we could also design discrete prompts for recommendation explanation generation.
As IDs are inherently different from words, one naive and straightforward way is to convert IDs into words, such as movie titles and item features.
We opt for the latter, and utilize features related to both the target user and the target item, since they represent the user's explicit preferences as well as the item's fine-grained attributes.
Moreover, these features could guide the model to talk about certain topics when generating explanations, such as ``room'' and ``location'' for hotel recommendations.

However, the conversion process from IDs into features may lose certain information, e.g., the identification role.
Specifically, it is not very likely to convert an ID back from some features.
For example, from the fact that Jerry loves cheese, we would not be able to certify that someone who enjoys eating cheese must be Jerry.
Moreover, prompts do not have to be text strictly.
They could be vectors, either randomly initialized or produced by another model.
This type of prompt is formally termed \textit{continuous/soft prompt}.
In a similar way, we can also input ID vectors to a pre-trained model for explanation generation.
Specifically, they are concatenated with the word vectors of an explanation before passing through the pre-trained model.
It is unnecessary to do so for the aforementioned discrete prompt, because discrete prompt is composed of words (i.e., features) and thus is consistent with the model.

A follow-up problem of continuous prompt is that the model is already trained, but the ID vectors are randomly initialized, so they are actually in different learning stages.
Recent study \cite{ICML19-convergence} finds out that such randomly initialized vectors could not be well optimized via stochastic gradient descent, and thus may lead to sub-optimal results.
To cope with the problem, we propose two training strategies.
The first strategy is called \textit{sequential tuning}, where we separate the training into two stages: fine-tune continuous prompts (i.e., ID vectors) with the model frozen, and then update the parameters of both.
The first stage would enable the continuous prompts to reach the same learning stage as the model, so that in the second stage they could be trained together.
Our second strategy named \textit{recommendation as regularization} is inspired by recent findings \cite{SIGIR16-LRPPM, TIST22-BPER, WWW20-DualPC} in explainable recommendation that the explanation performance could be improved by the recommendation task.
Indeed, the rating scores represent how much a user appreciates an item, which makes them an informative signal to the learning of explanation generation.
Hence, we also leverage rating prediction task to augment the explanation task, and test two typical recommendation models, including Matrix Factorization (MF) \cite{NIPS08-PMF} and Multi-Layer Perceptron (MLP).

We name our method PEPLER\footnote{Codes available at \url{https://github.com/lileipisces/PEPLER}}, which stands for ``PErsonalized Prompt Learning for Explainable Recommendation'', where personalization is reflected by the IDs, either implicitly in the discrete prompts or explicitly in the continuous prompts.
Without bells and whistles, our method consistently achieves the best performance against strong baselines (built on top of LSTM \cite{Neural97-LSTM}, GRU \cite{EMNLP14-GRU}, Transformer \cite{NIPS17-Transformer} or BERT \cite{NAACL19-BERT}) in terms of both text quality and explainability on three datasets.

In summary, our key contributions are:
\begin{itemize}
	\item We propose PEPLER that generates natural language explanations for recommendations by treating user and item IDs as prompts.
	To the best of our knowledge, we are the first to introduce prompt learning to the community of recommender systems.
	\item We propose two training strategies to bridge the gap between continuous prompts and the pre-trained model, in order to enhance the explanation generation performance.
	In a broader sense, this may inspire researchers on how to better tune pre-trained language models.
	\item We evaluate the generated explanations on not only text quality metrics (such as BLEU and ROUGE), but also metrics that particularly focus on explainability from the angle of item features.
	Extensive experiments show that our method consistently outperforms state-of-the-art baselines.
	\item Our work may shed light on a broader scope of natural language generation fields that also need personalization, e.g., personalized conversational systems.
	In addition, it may point out a way for pre-trained models to deal with multi-modal data, e.g., image and text in captioning systems.
\end{itemize}

In what follows, we first summarize related literature in section \ref{sec:related}, and then present our explanation generation method PEPLER in section \ref{sec:methodology}.
Experimental setup and results analysis are given in section \ref{sec:setup} and \ref{sec:results}, respectively.
We make a final conclusion and discuss future works in section \ref{sec:conclude}.

\section{Related Work} \label{sec:related}

\subsection{Explainable Recommendation}

Explainable recommendation \cite{FTIR20-Survey, Handbook15-Explanation} has been studied from two major perspectives: human-computer interaction and machine learning.
The former investigates how people perceive different styles of explanation \cite{IJHCS14-HCI, IUI17-HCI, chen2019user}, while the latter provides explanations by designing new explainable recommendation algorithms, to which our work is more related.
There exist various types of explanation style, such as pre-defined templates \cite{SIGIR14-EFM, CIKM21-Counterfactual, JIIS21-CAESAR}, item features \cite{CIKM15-TriRank, WWW18-TEM}, ranked text \cite{AAAI19-DER, WWW18-NARRE, SIGIR21-EXTRA}, image visualizations \cite{SIGIR19-VECF}, knowledge graph paths \cite{MDPI18-KG, SIGIR19-PGPR, SIGIR20-KGAT, CIKM20-CAFE}, and reasoning rules \cite{shi2020neural, chen2021neural, zhu2021faithfully}.
In this work, we focus on generating natural language explanations because they can be easily incorporated into different application scenarios, such as food recommender systems (e.g., Meituan\footnote{https://www.meituan.com/} \cite{CIKM20-Tip}) and conversational recommender systems \cite{IJCAI20-ECR, NeurIPS18-CRS, CIKM18-SAUR}.
However, previous works \cite{EARS19-HSS, IJCAI19-CAML, CIKM20-NETE, WSDM21-SAER} mostly rely on RNN, e.g., LSTM \cite{Neural97-LSTM} and GRU \cite{EMNLP14-GRU}, or unpretrained Transformer \cite{ACL21-PETER} for explanation generation, leaving the potentially more effective pre-trained models under-explored, which motivates this work.

\subsection{Transformer and Pre-trained Models}

Transformer \cite{NIPS17-Transformer} was first brought to the domain of machine translation with the architecture of encoder-decoder.
Later works \cite{ICLR18-Decoder, NAACL19-BERT} show that it remains effective, even when the encoder or the decoder is removed, reducing nearly half of model parameters.
Under the paradigm of pre-training plus fine-tuning, Transformer's effectiveness has been confirmed on a wide range of natural language understanding tasks \cite{18-GPT, NAACL19-BERT}, such as commonsense reasoning and question answering.
More recently, it has been shown that pre-trained Transformer is able to perform novel tasks on which it was not targeted during training, e.g., arithmetic, after increasing both the magnitude of model size and the volume of training corpus \cite{19-GPT2, NeurIPS20-GPT3}.
However, re-training such models may not be friendly to researchers who do not possess large amounts of computing resources.
Therefore, there emerges a new research direction: prompt learning \cite{CSUR22-Survey}, where researchers adapt their tasks to pre-trained models, without the need of modifying or re-training them.
Prompt learning has been successfully applied to many applications, such as domain adaptation \cite{TACL22-PADA}, text summarization \cite{ACL21-Prefix} and image captioning \cite{NeurIPS21-multimodal}, because it allows pre-trained models that contain rich world knowledge to perform different tasks with task-specific prompts.
In this work, we aim to provide users with high-quality recommendation explanations, so as to improve their experiences.
To this end, we explore recommendation-related prompts, including discrete prompt and continuous prompt.

\subsection{Personalized Natural Language Generation}

Personalization of natural language generation plays a vital role in a large spectrum of tasks, such as explainable recommendation \cite{EARS19-HSS, CIKM20-NETE, ACL21-PETER}, review summarization \cite{AAAI19-USN}, and dialog systems \cite{AAAI20-Chatbot, CIKM18-SAUR}.
In these tasks, user and item IDs are important identifiers for personalization.
Previous approaches typically adopt MLP to encode the IDs into a context vector, from which RNN can decode a word sequence.
This strategy can be found in many applications, such as review generation \cite{EACL17-Att2Seq, WWW19-MRG}, tip generation \cite{SIGIR17-NRT, WWW19-PATG} and explanation generation \cite{CIKM20-NETE, IJCAI19-CAML}.
However, it does not fit pre-trained models that were already trained on a massive amount of raw text.
Probably because a proper solution to deal with heterogeneous data (i.e., IDs and words) is yet to be invented, previous works with Transformer or pre-trained models for personalized natural language generation replace IDs with text segments, such as persona attributes \cite{AAAI20-Chatbot}, movie titles \cite{KDD20-Chatbot} and item features \cite{EMNLP19-ACMLM}, which is somewhat similar to our discrete prompt learning.
But besides this, we further investigate how to incorporate into pre-trained models continuous prompts (i.e., ID vectors), in order to retain as much information as possible.

\begin{table}
	\caption{Key notations and concepts.}
	\label{table:notation}
	\centering
	\begin{tabular}{l|l}
		\hline
		\textbf{Symbol} & \textbf{Description} \\
		\hline
		$\mathcal{T}$ & training set \\
		$\mathcal{U}$ & set of users \\
		$\mathcal{I}$ & set of items \\
		$\mathcal{V}$ & set of words \\
		$\mathcal{F}$ & set of features \\
		$\mathcal{E}$ & set of explanations \\
		\hline
		$\mathbf{U}$ & embeddings of users \\
		$\mathbf{I}$ & embeddings of items \\
		$\mathbf{u}$ & embedding of user $u$ \\
		$\mathbf{i}$ & embedding of item $i$ \\
		$\mathbf{c}$ & probability distribution over the vocabulary \\
		\hline
		$\mathbf{W}$ & weight matrix \\
		$\mathbf{w}, \mathbf{b}$ & weight vector \\
		$b$ & weight scalar \\
		$\mathbf{M}$ & attention masking matrix \\
		$\Theta$ & model parameters \\
		\hline
		$E$ & word sequence of an explanation \\	
		$d$, $d_{ff}$, $d_h$ & dimension of representation \\
		$m$ & number of attention heads \\
		$n$ & number of Transformer layers \\
		$z$ & number of MLP hidden layers \\
		\hline
		$\text{ReLU}(\cdot)$ & ReLU activation function \\
		$\sigma(\cdot)$ & sigmoid activation function \\
		$\text{softmax}(\cdot)$ & softmax function \\
		\hline
	\end{tabular}
\end{table}

\section{Methodology} \label{sec:methodology}

The goal of our explanation task is to generate a natural language sentence $\hat{E}_{u, i}$ for a given user-item pair $(u, i)$ to justify why $i$ is recommended to $u$.
The item $i$ could be predicted for the user $u$ by a recommendation model, e.g., matrix factorization \cite{NIPS08-PMF}, or resulted from his/her true behavior.
At both training and testing stages, only user $u$ and item $i$ are used as input for producing the explanation.
Hence, our proposed explanation generation approaches are compatible with any recommendation model, in which user and item IDs are indispensable.

In this section, we present the details of our methodology.
First, we briefly go through Transformer, pre-trained language models, and prompt learning.
Then, we introduce our proposed two methods for explanation generation, including discrete prompt learning and continuous prompt learning.
After that, we illustrate how an explanation is generated during the inference stage.
At last, we present two strategies for continuous prompt learning: sequential tuning, and recommendation as regularization.

Before introducing the technical details, we briefly explain the key terminology and notations.
A \textit{token} is a general term that can refer to \textit{user ID}, \textit{item ID}, \textit{word} and \textit{sub-word}.
An item \textit{feature} (e.g., ``room'') is also a \textit{word}, and thus can be seen as a \textit{token}.
A \textit{discrete prompt} is a word sequence, e.g., several item features, while a \textit{continuous prompt} is a sequence of vectors, e.g., user and item embeddings in this work.
Key notations and concepts are given in Table \ref{table:notation}.
We use italic upper-case to denote a sequence of tokens, e.g., $S$, and italic lower-case to indicate its composing units, e.g., $s$.
Meanwhile, a matrix is represented with bold upper-case, e.g., $\mathbf{S}$, and a vector is denoted as bold lower-case, e.g., $\mathbf{s}$, no matter whether they carry subscript or superscript or not.

\subsection{Transformer, Pre-trained Language Models and Prompt Learning}

To better demonstrate our work of PErsonalized Prompt Learning for Explainable Recommendation (PEPLER), we briefly go through Transformer and pre-trained language models that this work is built upon.
Transformer \cite{NIPS17-Transformer} consists of $n$ identical layers.
The $l$-th layer encodes the previous layer's output $\mathbf{S}_{l - 1}$ into $\mathbf{S}_l \in \mathbb{R}^{\left| S \right| \times d}$, where $l \in [1, n]$, $\left| S \right|$ is the length of the input token sequence, and $d$ denotes the dimension of token representations/embeddings.
Each layer is composed of two sub-layers: multi-head self-attention (MHSA) and position-wise feed-forward network (FFN).
The latter is a two-layer FFN with the ReLU activation function.
It performs linear transformations on the MHSA's output $\mathbf{O}_{l} \in \mathbb{R}^{\left| S \right| \times d}$, and converts $\mathbf{O}_{l}$ into $\mathbf{S}_l$,
\begin{equation}
	\mathbf{S}_{l} = \text{ReLU}(\mathbf{O}_{l} \mathbf{W}_{l, 1} + \mathbf{b}_{l, 1}) \mathbf{W}_{l, 2} + \mathbf{b}_{l, 2}
\end{equation}
where $\mathbf{W}_{l, 1} \in \mathbb{R}^{d \times d_{ff}}, \mathbf{b}_{l, 1} \in \mathbb{R}^{d_{ff}}, \mathbf{W}_{l, 2} \in \mathbb{R}^{d_{ff} \times d}, \mathbf{b}_{l, 2} \in \mathbb{R}^d$ are weight parameters.

The MHSA sub-layer aggregates $m$ attention heads, each of which is computed identically with the scaled dot-product attention (e.g., the $h$-th head in the $l$-th layer $\mathbf{A}_{l, h} \in \mathbb{R}^{\left| S \right| \times \frac{d}{m}}$).
Formally, the computation of this sub-layer is defined as follows:
\begin{equation}
	\begin{aligned}
		\mathbf{O}_{l} &= [\mathbf{A}_{l, 1}, ..., \mathbf{A}_{l, m}] \mathbf{W}_{l}^O \\
		\mathbf{A}_{l, h} & = \text{softmax} (\frac{\mathbf{Q}_{l, h} \mathbf{K}_{l, h}^\top}{\sqrt{d}} + \mathbf{M}) \mathbf{V}_{l, h} \\
		\mathbf{Q}_{l, h} & = \mathbf{S}_{l - 1} \mathbf{W}_{l, h}^Q, \mathbf{K}_{l, h} = \mathbf{S}_{l - 1} \mathbf{W}_{l, h}^K, \mathbf{V}_{l, h} = \mathbf{S}_{l - 1} \mathbf{W}_{l, h}^V \\
		\mathbf{M} & =
		\begin{cases}
			0, & \text{Allow to attend} \\
			- \infty, & \text{Prevent from attending}
		\end{cases}
	\end{aligned}
\end{equation}
where $[\cdot, \cdot]$ represents the concatenation of matrices/vectors, $\text{softmax} (\cdot)$ denotes the softmax function, $\mathbf{W}_{l}^O \in \mathbb{R}^{d \times d}$ and $\mathbf{W}_{l, h}^Q, \mathbf{W}_{l, h}^K, \mathbf{W}_{l, h}^V \in \mathbb{R}^{d \times \frac{d}{m}}$ are projection matrices to be learned, $\mathbf{S}_{l - 1} \in \mathbb{R}^{\left| S \right| \times d}$ is the $(l - 1)$-th layer's output, and $\mathbf{M} \in \mathbb{R}^{\left| S \right| \times \left| S \right|}$ is the attention masking matrix.

\begin{figure}
	\centering
	\includegraphics[scale=0.4]{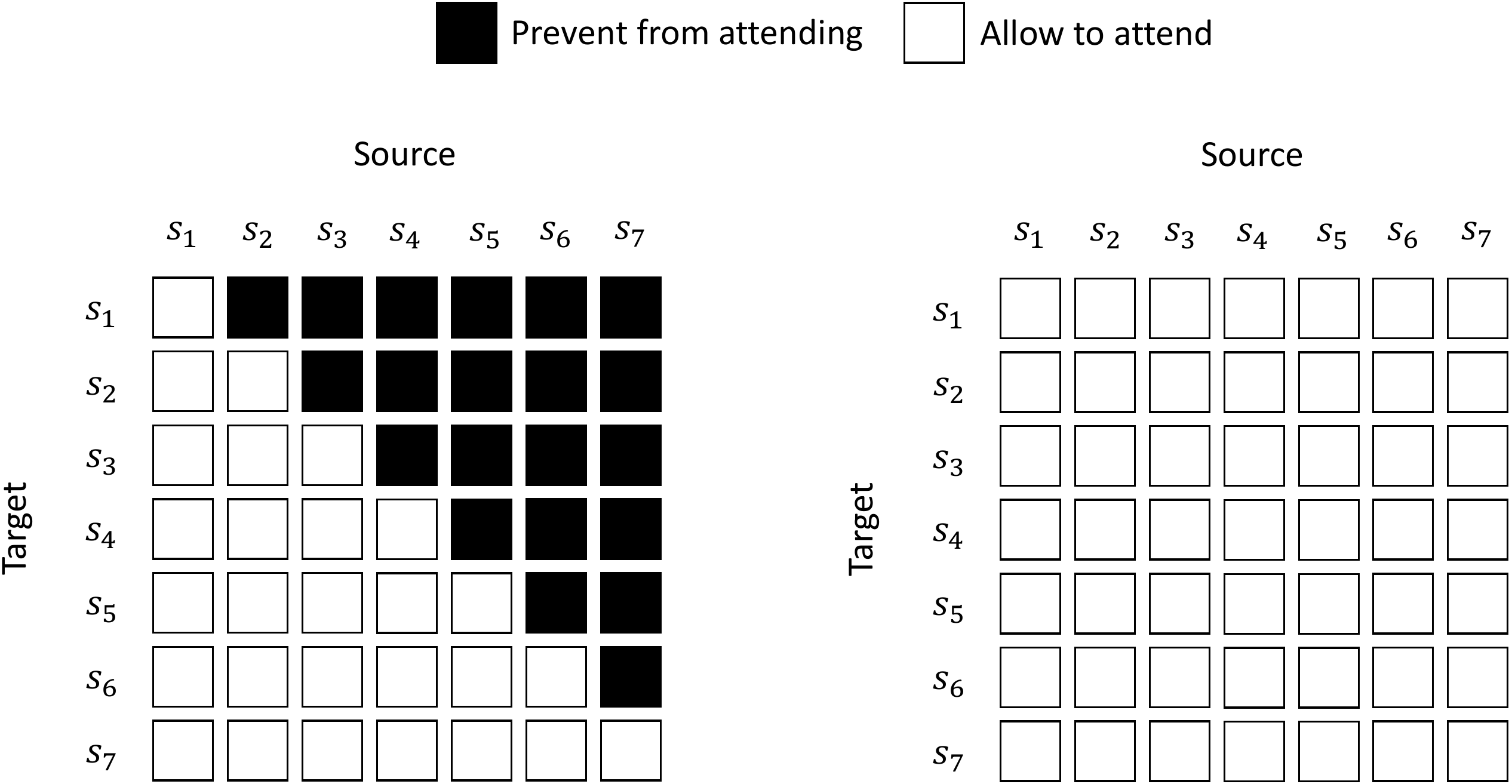}
	\caption{A comparison between left-to-right unidirectional attention masking (left) and bidirectional attention masking (right).}
	\label{fig:mask}
\end{figure}

Each element in $\mathbf{M}$ controls whether a token in the sequence can attend to another.
For example, in bidirectional language models such as BERT \cite{NAACL19-BERT}, $\mathbf{M}$ is a zero matrix that allows all tokens in the sequence to attend to each other.
Owing to the bidirectionality nature, this type of model is more suitable for natural language understanding tasks.
In the case of natural language generation, future tokens would be exposed to bidirectional language models, making them incapable of predicting these tokens.
As a comparison, left-to-right unidirectional language models, e.g., GPT \cite{18-GPT}, are particularly designed for natural language generation.
Specifically, in these models, the lower triangular part of $\mathbf{M}$ is set to $0$ and the remaining part $- \infty$, so as to allow each token to attend to past tokens (including itself), but prevent it from attending to future tokens.
A graphical comparison between the two types of attention masking mechanism is shown in Fig. \ref{fig:mask}.

With the two types of masking mechanism, there are also two corresponding pre-training objectives: cloze task, which is formally termed Masked Language Model (MLM) \cite{NAACL19-BERT}, for bidirectional language models, and auto-regressive generation for unidirectional language models.
Because our explanation generation task is closely related to the latter, we describe it in more details.
Specifically, given the output vectors $\mathbf{S}_n = [\mathbf{s}_{n, 1}, ..., \mathbf{s}_{n, \left| S \right|}]$ resulting from the last layer of Transformer, we pass them through a linear layer to obtain the probability distribution over all tokens in the dataset.
With the token probability distribution, we then make the next-token prediction based on preceding tokens, which can be achieved by minimizing the following negative log-likelihood:
\begin{equation}
	\mathcal{L} = \sum_t -\log p (s_t | s_{t - k}, ..., s_{t - 1}; \Theta_{LM})
\end{equation}
where $s_t$ is the next token to be predicted, $k$ denotes the size of the sliding context window, and $\Theta_{LM}$ represents all parameters in Transformer.

The pre-trained language models refer to those Transformers that have a great number of parameters (e.g., 1 billion) and were trained on a large volume of textual data (e.g., 100GB).
As a consequence, unlike small unpretrained Transformer \cite{ACL21-PETER}, it is less likely to do customized modifications on them.
In the meantime, re-training a large Transformer model would be unaffordable for most researchers who do not possess much computing resources.
Fortunately, there is a promising solution called \textit{prompt learning} \cite{CSUR22-Survey}, where different natural language processing tasks are adapted to a pre-trained language model so as to enable direct modeling of text.
In this way, the knowledge exhibited in the model could also be made good use of.

\begin{table}
	\caption{Prompt learning for typical natural language processing tasks \cite{CSUR22-Survey}. In the Template column, \underline{X} and \underline{Y} denote Input and Output, respectively. In our explanation generation task, the template words ``Explain the recommendation:'' are removed.}
	\label{table:prompt}
	\centering
	\begin{tabular}{llll}
		\hline
		\textbf{Task} & \textbf{Input (X)} & \textbf{Template} & \textbf{Output (Y)} \\
		\hline
		\multirow{3}{*}{Sentiment Classification} & \multirow{3}{*}{I love this book.} & \multirow{3}{*}{\underline{X} The book is \underline{Y}} & great \\
		&&& boring \\
		&&& ... \\ \hline
		\multirow{3}{*}{Text Summarization} & \multirow{3}{*}{The Omicron ...} & \multirow{3}{*}{\underline{X} TL;DR: \underline{Y}} & COVID-19 ... \\
		&&& Pandemic ... \\
		&&& ... \\ \hline
		%\multirow{4}{*}{Machine Translation} && \multirow{4}{*}{Chinese: \underline{X} English: \underline{Y}} & In me the ... \\
		%&\begin{CJK}{UTF8}{gbsn}心有猛虎，\end{CJK}&& The heart ... \\
		%&\begin{CJK}{UTF8}{gbsn}细嗅蔷薇。\end{CJK}&& Heart has ... \\
		%&&& ... \\ \hline
		\multirow{3}{*}{Machine Translation} && \multirow{3}{*}{French: \underline{X} English: \underline{Y}} & She tamed ... \\
		&Elle m\textquotesingle a apprivois\'e.\tablefootnote{It is an excerpt from a dialogue in a famous French novella \textit{The Little Prince}. The complete dialogue is ``Il y a une fleur. Je crois qu\textquotesingle elle m\textquotesingle a apprivois\'e'', meaning ``There is a flower. I think that she has tamed me''.}&& The flower ... \\
		&&& ... \\ \hline
		%https://microtop.ca/lepetitprince/chapitre21.html
		%https://www.angelfire.com/hi/littleprince/framechapter21.html
		\multirow{4}{*}{\textbf{Explanation Generation}} & \multirow{2}{*}{room location ...} & \underline{X} (Explain the & The room ... \\
		&& recommendation:) \underline{Y} & The breakfast ... \\ \cline{2-3}
		& X1: user123abc & \underline{X1} \underline{X2} (Explain the & The location ... \\
		& X2: item456def & recommendation:) \underline{Y} & ... \\ \hline
	\end{tabular}
\end{table}

Taking sentiment classification as an example, conventionally the prediction made by a model for a sample ``I love this book'' should be close to 1 (e.g., 0.97), indicating a positive sentiment.
In prompt learning, a template such as ``\underline{X} The book is \underline{Y}'' is constructed firstly.
Then, the input placeholder \underline{X} is filled in with a sample, e.g., ``\underline{I love this book.} The book is \underline{Y}'', which is termed \textit{prompt}.
With this, the model can be instructed to make a prediction at the output placeholder \underline{Y}, e.g., ``\underline{great}'' or ``\underline{boring}''.
At last, the prediction is mapped onto a sentiment, i.e., 1 or 0.
Clearly, there are two major steps that cost human efforts.
The first one is to manually design templates for different application scenarios, and to find the one that best fits a target application.
The second is the answer mapping stage, where a number of answer words need to be prepared in advance.

But it does not have to be so sophisticated for natural language generation tasks, whose input and output are both text \textit{per se}.
For example, the template for text summarization could simply be ``\underline{X} TL;DR: \underline{Y}''\footnote{``TL;DR'' stands for ``Too Long; Did/Do not Read''.}, and that for machine translation ``French: \underline{X} English: \underline{Y}''.
In a similar way, we could also define the template for explanation generation as ``\underline{X} Explain the recommendation: \underline{Y}''.
Although intuitively the template words may look useful, it was found that they could not always guide pre-trained language models to perform the specified task (e.g., ``summarize the table'') \cite{ACL21-Prefix}.
Moreover, our key focus is to automatically generate explanations for recommendations rather than manually constructing templates.
Therefore, we omit these template words, which gives us ``\underline{X} \underline{Y}'' and ``\underline{X1} \underline{X2} \underline{Y}''.
A comparison of prompt learning between the aforementioned tasks is given in Table \ref{table:prompt}.
In the following, we describe our proposed two methods for explainable recommendation: discrete prompt learning and continuous prompt learning.

\subsection{Discrete Prompt Learning} \label{sec:discrete}

\begin{figure}
	\centering
	\includegraphics[scale=0.4]{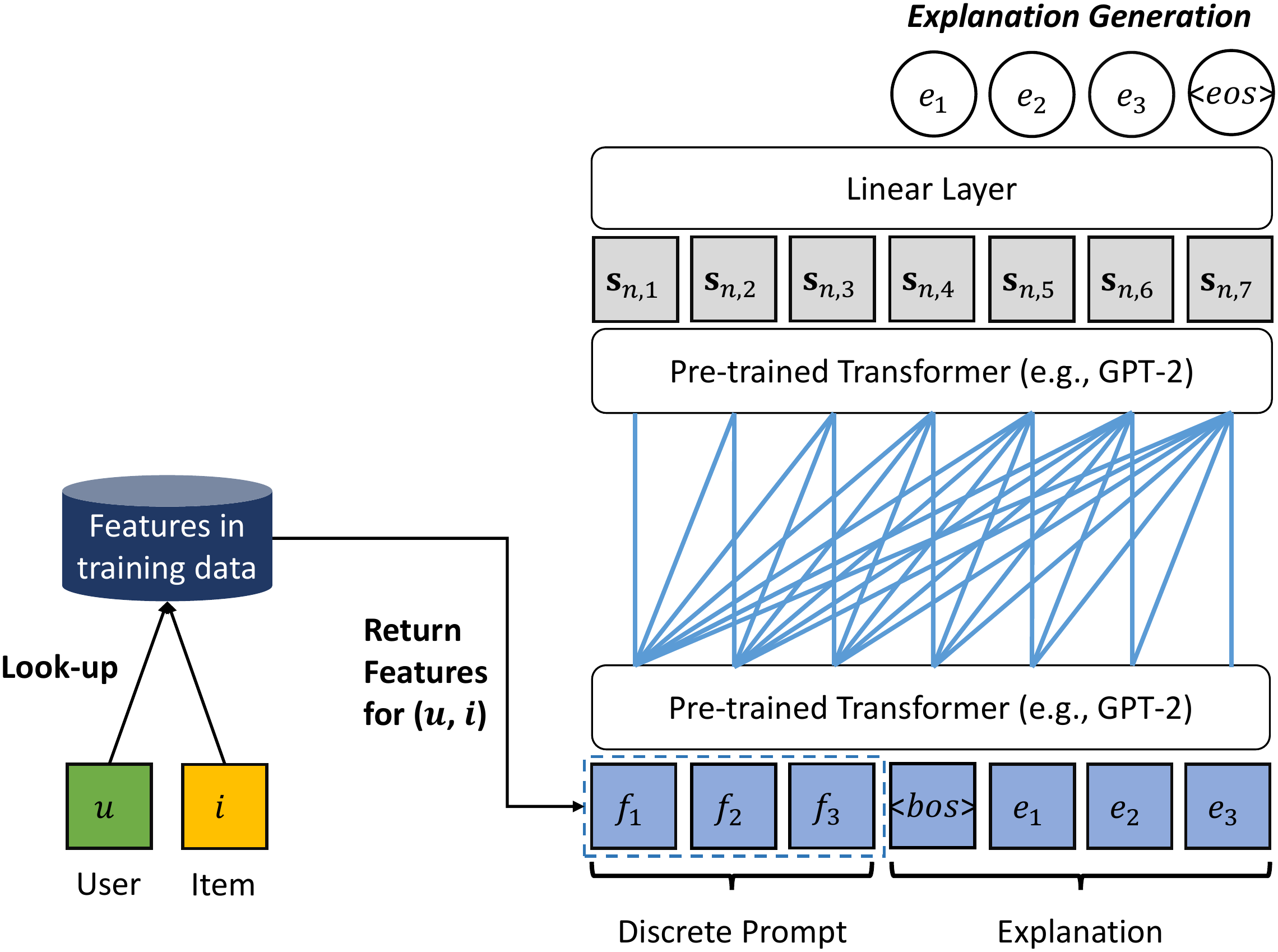}
	\caption{Our proposed method PEPLER-D that utilizes item features as discrete prompt for explanation generation.}
	\label{fig:discrete}
\end{figure}

Pre-trained language models, such as BERT \cite{NAACL19-BERT} and GPT-2 \cite{19-GPT2}, were trained on a large amount of words, which are inherently in a different semantic space as ID tokens, but IDs (e.g., user ID) are indispensable in recommender systems.
To resolve this issue, a straightforward way is to find some domain-specific words to represent the IDs, such as movie titles and item features (e.g., ``bedroom'' for hotel recommendation).
In this way, a pre-trained model can be prompted to generate recommendation-specific text.
In this work, we explore item features for recommendation explanation generation, and denote the proposed approach as PEPLER-D, where ``D'' stands for ``discrete prompt learning''.
A graphical illustration of PEPLER-D is shown in Fig. \ref{fig:discrete}.

From the training set, we can obtain all the item features $F_u$ (or $F_i$) associated with a user $u$ (or an item $i$).
Suppose $F_u = \{\text{gym}, \text{bathroom}, \text{breakfast}\}$, and $F_i = \{\text{gym}, \text{breakfast}, \text{subway}, \text{Wi-Fi}\}$.
For efficiency, we set the discrete prompt to a fixed size (e.g., 4 in this toy example), which is a common strategy in recommender systems.
Under this setting, we need to ensure that the discrete prompt contains as many informative item features as possible, so as to allow the pre-trained model to generate high-quality explanations.
For each user-item pair $(u, i)$, the features in $F_u \cap F_i = \{\text{gym}, \text{breakfast}\}$ are more informative because they are related to both user $u$ and item $i$.
However, when $F_u \cap F_i$ is small and does not reach the size of the discrete prompt, we also take the other features in $(F_u \cup F_i) / (F_u \cap F_i) = \{\text{bathroom}, \text{subway}, \text{Wi-Fi}\}$ into consideration.
Though less informative, they are at least associated with either user $u$ or item $i$.
Then, the discrete prompt for the user-item pair is defined as:
\begin{equation}
	F_{u, i} = [(F_u \cap F_i), (F_u \cup F_i) / (F_u \cap F_i)]
\end{equation}
Because the prompt size in the example is fixed to 4, we only use $[\text{gym}, \text{breakfast}, \text{bathroom}, \text{subway}]$ in $F_{u, i}$ for explanation generation, and drop the other item features.

During the training stage, the input sequence to the pre-trained model can be represented as $S = [f_1, \cdots, f_{\left| F_{u, i} \right|}, e_1, \cdots, e_{\left| E_{u, i} \right|}]$, where $f_1, \cdots, f_{\left| F_{u, i} \right|}$ are the discrete prompt consisting of features, $e_1, \cdots, e_{\left| E_{u, i} \right|}$ are the explanation's word sequence, and $\left| F_{u, i} \right|$ and $\left| E_{u, i} \right|$ denote the number of features and explanation words, respectively.
Because all the tokens in sequence $S$ are of the same type, i.e., words, we can perform embedding look-up once for them all, which gives the sequence's token representation $[\mathbf{f}_1, \cdots, \mathbf{f}_{\left| F_{u, i} \right|}, \mathbf{e}_1, \cdots, \mathbf{e}_{\left| E_{u, i} \right|}]$.
The input representation of the sequence to the model is the addition of the token representation, and the positional representation $[\mathbf{p}_1, \cdots, \mathbf{p}_{\left| S \right|}]$ that encodes the position of each token in the sequence.
We denote the input representation as $\mathbf{S}_0 = [\mathbf{s}_{0, 1}, \cdots, \mathbf{s}_{0, \left| S \right|}]$, where $\left| S \right|$ is the length of the sequence.

After passing $\mathbf{S}_0$ through pre-trained Transformer, we obtain the sequence's final representation $\mathbf{S}_n = [\mathbf{s}_{n, 1}, \cdots, \mathbf{s}_{n, \left| S \right|}]$.
Then, we apply a linear layer to each token's final representation to map it onto a $\left| \mathcal{V} \right|$-sized vector.
As an example, $\mathbf{s}_{n, t}$ becomes $\mathbf{c}_t$ after passing through this layer:
\begin{equation}
	\mathbf{c}_t = \text{softmax} (\mathbf{W}^v \mathbf{s}_{n, t} + \mathbf{b}^v)
\end{equation}
where $\mathbf{W}^v \in \mathbb{R}^{\left| \mathcal{V} \right| \times d}$ and $\mathbf{b}^v \in \mathbb{R}^{\left| \mathcal{V} \right|}$ are weight parameters, and $\text{softmax}(\cdot)$ is the softmax function.
The vector $\mathbf{c}_t$ represents the probability distribution over the vocabulary $\mathcal{V}$.
For model learning, we adopt negative log-likelihood (NLL) as the loss function, and compute the mean of user-item pairs in the training set:
\begin{equation}
	\mathcal{L}_D = \frac{1} {\left| \mathcal{T} \right|} \sum_{(u, i) \in \mathcal{T}} \frac{1}{\left| E_{u, i} \right|} \sum_{t = 1}^{\left| E_{u, i} \right|} - \log c_{\left| F_{u, i} \right| + t}^{e_t}
	\label{eq:discrete}
\end{equation}
where the probability $c_t^{e_t}$ is offset by $\left| F_{u, i} \right|$ positions because the explanation is placed at the end of the sequence.

\subsection{Continuous Prompt Learning} \label{sec:continuous}

We have shown that it is feasible to use item features as discrete prompt to a pre-trained model for explanation generation.
However, the conversion from IDs to words (i.e., features) may lose some important information of IDs.
Taking the identification role of IDs as an example, it is nearly impossible to convert the features back into IDs.
Meanwhile, prompts do not necessarily have to be words or even readable.
They can be vector representations, either produced by other models or randomly initialized.
This type of human-incomprehensible prompts are formally termed \textit{continuous/soft prompt}.
Thus, ID vectors could also be directly used as continuous prompts to generate recommendation explanations.
Next, we show how to encode the two types of ID $u$ and $i$ into vector representations.

\begin{figure}
	\centering
	\includegraphics[scale=0.4]{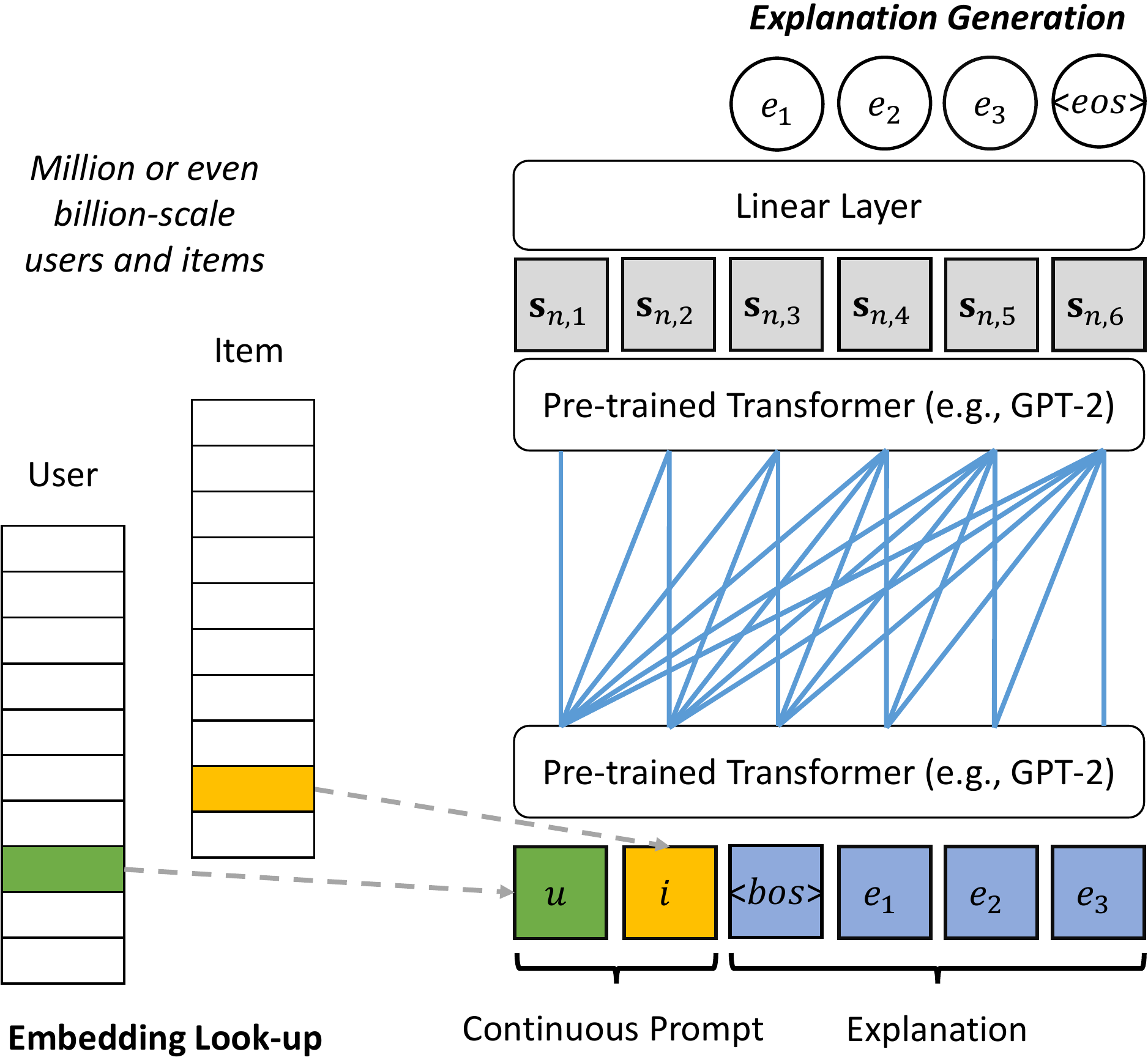}
	\caption{Our proposed method PEPLER that treats user and item IDs as continuous prompt for explanation generation.}
	\label{fig:continuous}
\end{figure}

Conceptually, the input sequence can be represented as $S = [u, i, e_1, \cdots, e_{\left| E_{u, i} \right|}]$, as shown in Fig. \ref{fig:continuous}.
Intuitively, one may regard the IDs as special word tokens, and add them to the pre-trained model's vocabulary $\mathcal{V}$.
However, there could be millions or even billions of users and items in recommender systems (e.g., in e-commerce).
When generating explanations, predicting a word out of the huge amount of IDs would be time-consuming.
Therefore, we do not add the IDs to $\mathcal{V}$, but instead treat them as two additional types of tokens.
Specifically, we prepare two sets of token embeddings: $\mathbf{U} \in \mathbb{R}^{\left| \mathcal{U} \right| \times d}$ and $\mathbf{I} \in \mathbb{R}^{\left| \mathcal{I} \right| \times d}$, where $\vert \mathcal{U} \vert$ and $\vert \mathcal{I} \vert$ represent the number of users and items in a dataset, respectively.
Then, a user $u$'s vector representation can be retrieved via:
\begin{equation}
	\mathbf{u} = \mathbf{U}^\top {\rm g} (u)
\end{equation}
where ${\rm g} (u) \in \{0, 1\}^{\vert \mathcal{U} \vert}$ denotes a one-hot vector, whose non-zero element corresponds to the position that user $u$'s vector locates in $\mathbf{U}$.
In a similar way, we can obtain $\mathbf{i}$ from $\mathbf{I}$ for item $i$.
Notice that, the embeddings $\mathbf{U}$ and $\mathbf{I}$ are randomly initialized, but will be updated by back-propagation during the training process.
Then, the sequence's token representation can be denoted as $[\mathbf{u}, \mathbf{i}, \mathbf{e}_1, \cdots, \mathbf{e}_{\left| E_{u, i} \right|}]$.

The follow-up steps are identical to discrete prompt learning in Section \ref{sec:discrete}: perform addition for token representation and positional representation to obtain $\mathbf{S}_0 = [\mathbf{s}_{0, 1}, \cdots, \mathbf{s}_{0, \left| S \right|}]$, pass $\mathbf{S}_0$ through pre-trained Transformer for producing $\mathbf{S}_n = [\mathbf{s}_{n, 1}, \cdots, \mathbf{s}_{n, \left| S \right|}]$, apply a linear layer with softmax function to each token's final representation $\mathbf{s}_{n, t}$ for next-word prediction, and employ NLL loss function on the word probability distribution $\mathbf{c}_t$:
\begin{equation}
	\mathcal{L}_C = \frac{1} {\left| \mathcal{T} \right|} \sum_{(u, i) \in \mathcal{T}} \frac{1}{\left| E_{u, i} \right|} \sum_{t = 1}^{\left| E_{u, i} \right|} - \log c_{2 + t}^{e_t}
	\label{eq:continuous}
\end{equation}
where $c_t^{e_t}$ is offset by 2 positions (i.e., user ID and item ID), which is slightly different multiple positions of features in Eq. \eqref{eq:discrete}.

\subsection{Explanation Generation}

During the inference stage, our goal is to instruct the model to generate a word sequence $E^*$, which has the maximum log-likelihood, as explanation.
\begin{equation}
	E^* = \mathop{\arg\max}_{E \in \hat{\mathcal{E}}} \sum_t^{\left| E \right|} \log c_{\left| prompt \right| + t}^{e_t}
\end{equation}
where $\hat{\mathcal{E}}$ is the set of all generated word sequences, and $\left| prompt \right|$ denotes the prompt's length, i.e., 2 for $[u, i]$ and $\left| F_{u, i} \right|$ for $F_{u, i}$.

There are various methods to find the sequence $E^*$, such as greedy decoding and beam search.
Since it is not our key focus to develop searching algorithms, we adopt the simple greedy decoding, which treats the word with the largest probability as the prediction at each step.
More precisely, along with the prompt $u$ and $i$ (or $F_{u, i}$), we first feed the model a special begin-of-sequence token $<$\textit{bos}$>$.
From the resulting word probability distribution $\mathbf{c}_{<bos>}$, we can select the highest probability word as prediction.
Then, we concatenate this predicted word at the end of the sequence to form a new input sequence for generating another word.
We do this repeatedly until the model produces a special end-of-sequence token $<$\textit{eos}$>$, or the generated explanation reaches a pre-defined length.

\subsection{Sequential Tuning Strategy} \label{sec:seq}

In the case of discrete prompt learning, the prompts are features, which are of the same type as words that pre-trained language models were trained on.
As a result, no additional model parameters are introduced, so we can simply optimize Eq. \eqref{eq:discrete} with the following objective function:
\begin{equation}
	\mathcal{J} = \min_{\Theta_{LM}} \mathcal{L}_D
\end{equation}
where $\Theta_{LM}$ denotes all the trainable parameters in the pre-trained language model.

However, in the case of continuous prompt learning, we introduced additional prompt parameters, i.e., two sets of embeddings for users and items.
Therefore, the model parameters $\Theta$ to be updated include pre-trained language model parameters $\Theta_{LM}$ and prompt parameters $\Theta_{P}$.
Obviously, the two types of parameters are in different learning stages, since the former are already trained from a large amount of textual data, while the latter are randomly initialized.
For example, it is easy to distinguish one word from another with the embeddings from $\Theta_{LM}$, e.g., ``hotel'' and ``room'', but it may not be that distinguishable for two users with random embeddings from $\Theta_{P}$, such as ``Tom'' and ``Jerry''.
Also, previous study \cite{ICML19-convergence} shows that randomly initialized parameters could only be updated in a small neighborhood with stochastic gradient descent (SGD).
Hence, how to effectively bridge the two types of parameters becomes a critical issue.

\begin{table}
	\caption{Different strategies for tuning pre-trained language models \cite{CSUR22-Survey}. ``Para.'' stands for parameters. ``N/A'' means that there is no prompt, while ``None'' indicates that the prompts do not have additional parameters.}
	\label{table:training}
	\centering
	\begin{tabular}{p{6cm}lll}
		\hline
		\textbf{Strategy} & \textbf{LM Para.} & \textbf{Prompt Para.} & \textbf{Typical Example} \\
		\hline
		Promptless Fine-tuning & Tuned & N/A & BERT \cite{NAACL19-BERT} \\
		Tuning-free Prompting & Frozen & None & GPT-3 \cite{NeurIPS20-GPT3} \\
		Fixed-LM Prompt Tuning & Frozen & Tuned & Prefix-Tuning \cite{ACL21-Prefix} \\
		Fixed-prompt LM Tuning & Tuned & None & Our PEPLER-D \\
		Prompt+LM Fine-tuning & Tuned & Tuned & P-Tuning \cite{21-PTuning} \\
		\hline
		\textbf{Sequential Tuning: Fixed-LM Prompt Tuning $\rightarrow$ Prompt+LM Fine-tuning} & \multirow{2}{*}{Tuned} & \multirow{2}{*}{Tuned Twice} & \multirow{2}{*}{Our PEPLER} \\
		\hline
	\end{tabular}
\end{table}

To tackle this problem, we propose a sequential tuning strategy.
Specifically, we first freeze the language model parameters $\Theta_{LM}$, and optimize the prompt parameters $\Theta_{P}$ with Eq. \eqref{eq:continuous}.
Once $\Theta_{P}$ has converged, we fine-tune all the model parameters (i.e., $\Theta_{LM}$ and $\Theta_{P}$) with Eq. \eqref{eq:continuous} again.
This two-step procedure can be demonstrated with the following formula:
\begin{equation}
	\mathcal{J} = \min_{\Theta_{P}} \mathcal{L}_C \xrightarrow[]{\text{followed by}} \mathcal{J} = \min_{\Theta = \{\Theta_{LM}, \Theta_{P}\}} \mathcal{L}_C
\end{equation}

In fact, our sequential tuning strategy is a combination of two typical tuning strategies \cite{CSUR22-Survey}: Fixed-LM Prompt Tuning and Prompt+LM Fine-tuning (see Table \ref{table:training}).
In section \ref{sec:sequential}, we conduct an effect comparison to prove that this strategy is indeed more useful than either of them.
We omit the other three strategies, i.e., Promptless Fine-tuning, Tuning-free Prompting and Fixed-prompt LM Tuning.
The first is usually used in pre-training plus fine-tuning paradigm, and the second is particularly suitable for zero-shot learning scenario, so they are not applicable to our methods.
The last one is adopted in our PEPLER-D.

\subsection{Recommendation as Regularization} \label{sec:reg}

\begin{figure}
	\centering
	\includegraphics[scale=0.4]{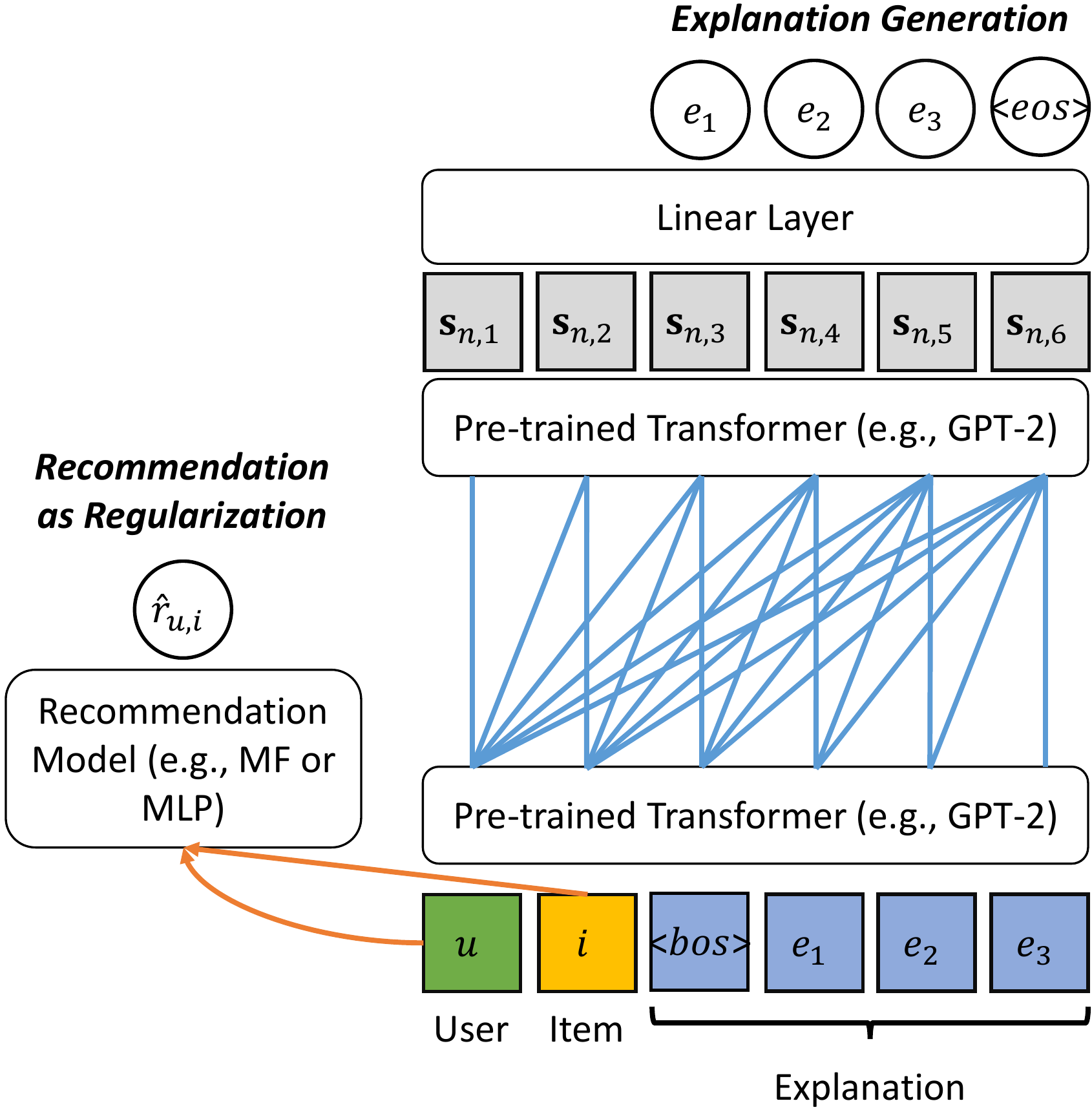}
	\caption{Our proposed method PEPLER that regards the rating prediction task as a type of regularization for better learning of the explanation generation task.}
	\label{fig:regularization}
\end{figure}

To bridge the aforementioned gap between pre-trained language models and continuous prompts, we come up with another approach: regularizing the learning of explanation generation via an additional rating prediction task (see Fig. \ref{fig:regularization}).
The intuition behind this idea is that each rating score $r_{u, i}$ was assigned by a user $u$ to an item $i$, so it to some extent captures the relation between this user-item pair.
Hence, the ratings could be used to better learn the continuous prompts.
Moreover, recent studies find out that the two task of recommendation and an additional task (such as feature ranking \cite{SIGIR16-LRPPM}, explanation ranking \cite{TIST22-BPER} and review generation \cite{WWW20-DualPC}) could help the learning of each other.
Inspired by this, we propose to leverage recommendation task to help the learning of explanation generation.
Since there is a great number of off-the-shelf recommendation models and our key focus is on explanation generation, we adopt and test two typical recommendation models: Matrix Factorization (MF) \cite{NIPS08-PMF} and Multi-Layer Perceptron (MLP) \cite{SIGIR17-NRT}.

Specifically, for MF the rating score $\hat{r}_{u, i}$ is resulted from the dot product of the target user and item's representations $\mathbf{u}$ and $\mathbf{i}$:
\begin{equation}
	\hat{r}_{u, i} = \mathbf{u}^\top \mathbf{i}
\end{equation}
Because the two types of representations are already available, this operation does not introduce additional model parameters.
In the case of MLP with $z$ hidden layers, the rating score is computed as follows:
\begin{equation}
	\left\{
	\begin{array}{lr}
		\mathbf{a}_0 = \sigma(\mathbf{W}_0 [\mathbf{u}, \mathbf{i}] + \mathbf{b}_0) & \\
		\mathbf{a}_1 = \sigma(\mathbf{W}_1 \mathbf{a}_0 + \mathbf{b}_1) & \\
		\dots\dots & \text{and}~~~\hat{r}_{u, i} = \mathbf{w}^\top \mathbf{a}_z + b \\
		\mathbf{a}_z = \sigma(\mathbf{W}_z \mathbf{a}_{z - 1} + \mathbf{b}_z) &
	\end{array}  
	\right.
\end{equation}
where $\mathbf{W}_0 \in \mathbb{R}^{d_{h} \times 2d}, \mathbf{b}_0 \in \mathbb{R}^{d_{h}}, \mathbf{W}_\ast \in \mathbb{R}^{d_{h} \times d_{h}}, \mathbf{b}_\ast \in \mathbb{R}^{d_{h}}, \mathbf{w} \in \mathbb{R}^{d_{h}}, b \in \mathbb{R}$ are additional parameters for the recommendation task, and $\sigma(\cdot)$ denotes the sigmoid function.
For both MF and MLP, mean square error is adopted as the loss function:
\begin{equation}
	\mathcal{L}_R = \frac{1} {\left| \mathcal{T} \right|} \sum_{(u, i) \in \mathcal{T}} (r_{u, i} - \hat{r}_{u, i})^2
\end{equation}
where $r_{u, i}$ is the ground-truth rating that user $u$ assigned to item $i$.

Then, the two tasks can be integrated into a multi-task learning framework with the following objective function:
\begin{equation}
	\mathcal{J} = \min_{\Theta = \{\Theta_{LM}, \Theta_{P}, \Theta_{REC}\}} (\mathcal{L}_C + \lambda \mathcal{L}_R)
\end{equation}
where the model parameters $\Theta$ consist of pre-trained language model parameters $\Theta_{LM}$, continuous prompt parameters $\Theta_{P}$ (i.e., user and item representations) and recommendation model parameters $\Theta_{REC}$ ($\emptyset$ for MF).
Since the recommendation task is used as a regularization term, we can adjust the regularization coefficient $\lambda$ to control the learning of the explanation generation task.

\section{Experimental Setup} \label{sec:setup}

\subsection{Datasets}

For experimentation, we adopt three publicly available explainable recommendation datasets, and their data splits \cite{CIKM20-NETE}.
During the splitting process, each dataset is randomly divided into training, validation and testing sets with ratio 8:1:1 for 5 times, and the training set holds at least one record for each user and each item.
The three datasets are from TripAdvisor\footnote{https://www.tripadvisor.com} (hotel), Amazon\footnote{http://jmcauley.ucsd.edu/data/amazon} (movies \& TV) and Yelp\footnote{https://www.yelp.com/dataset/challenge} (restaurant), respectively.
Each record in the datasets is comprised of a user ID, an item ID, a rating in the scale of 1 to 5, an explanation and an item feature.
The explanations are sentences extracted from user reviews.
Each explanation contains at least one item feature, such as ``bedroom'' and ``breakfast'', which ensures the explanation quality.
Statistics of the datasets are shown in Table \ref{table:dataset}.
We can see that Yelp is much larger than the other two in terms of size, making it closer to the real-world situation where there are millions of users and items.

\begin{table}
	\caption{Statistics of the three datasets.}
	\label{table:dataset}
	\centering
	\begin{tabular}{l|r|r|r}
		\hline
		& \textbf{TripAdvisor} & \textbf{Amazon} & \textbf{Yelp} \\
		\hline
		\#users & 9,765 & 7,506 & 27,147 \\
		\#items & 6,280 & 7,360 & 20,266 \\
		\#records & 320,023 & 441,783 & 1,293,247 \\
		\#features & 5,069 & 5,399 & 7,340 \\
		\#records / user & 32.77 &58.86&47.64 \\
		\#records / item & 50.96 &60.02&63.81 \\
		\#words / explanation & 13.01 &14.14&12.32 \\
		\hline
	\end{tabular}	
\end{table}

\subsection{Evaluation Metrics}

To evaluate explanation performance, we measure the generated explanations from two main perspectives: text quality and explainability.
For the former, we adopt \textbf{BLEU} \cite{ACL02-BLEU} in machine translation and \textbf{ROUGE} \cite{TS04-ROUGE} in text summarization, and report BLEU-1 and BLEU-4, and Precision, Recall and F1 of ROUGE-1 and ROUGE-2.
Notice that, BLEU is a precision-oriented metric, while ROUGE is a recall-oriented metric.
Though being widely used, BLUE and ROUGE are not flawless.
For example, it is difficult for them to detect the problem of identical sentences, i.e., many explanations for different user-item pairs are exactly the same for some methods, as shown in our experiments.
Treating these identical sentences as explanations is less appropriate, because they are less likely to well explain the special property of different recommendations.
To quantitatively measure this, we adopt \textbf{USR} that computes the Unique Sentence Ratio of generated explanations \cite{CIKM20-NETE}:
\begin{equation}
	USR = \frac{\left| \mathcal{E} \right|}{N}
\end{equation}
where $\mathcal{E}$ represents the set of unique sentences generated by a model, and $N$ is the total number of testing samples.
Note that, $\mathcal{E}$ only holds one of the exactly matched explanations.

Moreover, text quality is not equal to explainbility.
In the case of explainable recommendation, users may value more an explanation that justifies a recommendation's advantage on certain item features \cite{CIKM20-NETE, EARS19-HSS}.
To this end, we adopt the other three metrics proposed in \cite{CIKM20-NETE}: Feature Matching Ratio (\textbf{FMR}), Feature Coverage Ratio (\textbf{FCR}) and Feature Diversity (\textbf{DIV}).

FMR measures whether a generated explanation contains the feature in the ground-truth text.
Formally, it is defined as follows:
\begin{equation}
	FMR = \frac 1 N \sum_{u, i} \delta (f_{u, i} \in \hat{E}_{u, i})
\end{equation}
where $\hat{E}_{u, i}$ is the generated explanation for the user-item pair, $f_{u, i}$ is the feature in the ground-truth, and $\delta (x) = 1$ when $x$ is true, or $\delta(x) = 0$ otherwise.

FCR is computed as the number of distinct features contained in all the generated explanations, divided by the total number of features in the whole dataset:
\begin{equation}
	FCR = \frac {N_g} {\left| \mathcal{F} \right|}
\end{equation}
where $\mathcal{F}$ is the collection of unique features in ground-truth explanations, and $N_g$ denotes the amount of distinct features appeared in the generated explanations.

DIV measures the diversity of features between all generated explanations.
The intuition is that explanations are expected to discuss different features in accordance with the given user-item pairs.
Hence, it computes the intersection of features between any two generated explanations:
\begin{equation}
	DIV = \frac 2 {N \times (N - 1)} \sum_{u, u', i, i'} \left| \hat{\mathcal{F}}_{u, i} \cap \hat{\mathcal{F}}_{u', i'} \right|
\end{equation}
where $\hat{\mathcal{F}}_{u, i}$ and $\hat{\mathcal{F}}_{u', i'}$ represent two feature sets contained in two generated explanations, and $\left| \cdot \right|$ denotes the number of features in the resulting set.

For DIV, the lower, the better, while it is opposite for the rest of metrics.

\subsection{Compared Methods} \label{sec:methods}

We introduce four state-of-the-art baselines, which are based on representative language models, including BERT \cite{NAACL19-BERT}, Transformer \cite{NIPS17-Transformer}, GRU \cite{EMNLP14-GRU} and LSTM \cite{Neural97-LSTM}, respectively.
For these baselines, their whole model parameters are trained all together.
We divide them into two groups, depending on whether IDs are directly used or not.

We first compare our PEPLER-D with the following method, because both of them do not directly make use of IDs but instead map IDs onto item features.
\begin{itemize}
	\item \textbf{Aspect Conditional Masked Language Model (ACMLM)} \cite{EMNLP19-ACMLM} is a fine-tuned BERT \cite{NAACL19-BERT}, where an attention layer is introduced to encode the features for both the user and the item.
	By predicting masked tokens, this model can produce diverse sentences.
\end{itemize}

Then, we make comparison with the following three methods for our PEPLER, since they all leverage only user and item IDs to generate explanations.
\begin{itemize}
	\item \textbf{Neural Rating and Tips generation (NRT)} \cite{SIGIR17-NRT} can predict a rating and generate a tip simultaneously based on user and item IDs.
	The generation component is a GRU \cite{EMNLP14-GRU}.
	We take the explanations in the datasets as tips.
	Moreover, we find that the model's problem of generating identical sentences (as reported in \cite{CIKM20-NETE}) is caused by the L2 regularization in its original design.
	For fair comparison, we removed it.
	\item \textbf{Attribute-to-Sequence (Att2Seq)} \cite{EACL17-Att2Seq} is a review generation approach with a two-layer LSTM \cite{Neural97-LSTM}.
	We take the explanations as reviews.
	This model has an attention module, but we find that it makes the generated content unreadable.
	To be fair, we removed it as well.
	\item \textbf{PErsonalized Transformer for Explainable Recommendation (PETER)} \cite{ACL21-PETER} is a small unpretrained Transformer \cite{NIPS17-Transformer} particularly designed for explanation generation.
	To bridge the gap between IDs and words, an additional task named ``context prediction'' is introduced.
	This model can also make recommendations.
\end{itemize}

We conducted a user survey in NETE \cite{CIKM20-NETE, WWW20-NETE} and showed that the explanations generated by NETE were perceived useful by participants.
Moreover, the explanation quality of PETER \cite{ACL21-PETER} is much better than that of NETE on the same automatic evaluation metrics.
Hence, as long as the explanations produced by our new approach in this work are of even better quality than PETER on the same evaluation metrics, they shall be useful to real users as well.
This is evidenced by \cite{wen2022expscore} that users' perception towards machine-generated explanations are highly correlated with the factors of relevance, repetition and feature appearance, which correspond to BLEU/ROUGE, USR and FMR in this work.

\subsection{Implementation Details}

We train each model on the training set, tune the hyper-parameters on the validation set, and report the performance on the testing set.
The results are averaged on the 5 data splits.
We adopt the code of ACMLM, and implement the other baselines (i.e., NRT, Att2Seq and PETER) by ourselves.
For our models PEPLER and PEPLER-D, we implement them in Python\footnote{https://www.python.org} with PyTorch\footnote{https://pytorch.org}, and load pre-trained GPT-2 \cite{19-GPT2} from huggingface\footnote{https://huggingface.co/gpt2} as their backbone.
GPT-2 uses Byte Pair Encoding (BPE) \cite{ACL16-BPE} for vocabulary construction.
This technique could effectively mitigate Out-Of-Vocabulary (OOV) problem by encoding rare words into multiple sub-word units.
For example, the word ``restaurant'' is encoded into three sub-words ``rest'', ``aur'' and ``ant'', while the word ``room'' is still ``room''.
In total, there are 50,257 BPE tokens in GPT-2.
For fair comparison, we apply BPE to all the models, and set the length of explanations to 20 BPE tokens.
For our model PEPLER-D, the number of input features is also set to 20 BPE tokens.
We reuse the other default settings of the baselines.

The size of embeddings/representations $d$ in GPT-2 is 768.
We optimize our models PEPLER and PEPLER-D with AdamW \cite{ICLR19-AdamW}, and set batch size to 128.
The learning rate is set to 0.001 for PEPLER, and 0.0001 for PEPLER-D.
At each epoch, we save the model if it achieves the lowest loss on the validation set.
When the loss does not decrease for 5 times, we stop training and load the saved model for prediction.
In the case of recommendation as regularization in PEPLER, the number of hidden layers $z$ in MLP is set to 2, and the dimension of hidden layers $d_h$ 400.
We search the regularization coefficient $\lambda$ from $[10^{-5}, 10^{-4}, ..., 10^3]$.

\begin{table}[!tbp]
	\caption{Performance comparison of explanation generation methods in terms of Explainability and Text Quality on three datasets. The methods are divided into two groups according to whether IDs are directly used or not. PEPLER employs the default sequential tuning strategy, while the other two variants use recommendation as regularization with MLP and MF, respectively. B1 and B4 stand for BLEU-1 and BLEU-4. R1-P, R1-R, R1-F, R2-P, R2-R and R2-F denote Precision, Recall and F1 of ROUGE-1 and ROUGE-2. BLEU and ROUGE are percentage values (\% symbol omitted for table clarity), while the others are absolute values. The best performing values are boldfaced, and ** and * indicate the statistical significance over the best baseline for $p < 0.01$ and $p < 0.05$ via Student's t-test, respectively.}
	\label{table:explanation}
	\centering
	\setlength{\tabcolsep}{3.2pt}
	\footnotesize
	\begin{tabular}{r|llll|llllllll}
		\hline
		\multirow{2}{*}{} & \multicolumn{4}{c|}{\textbf{Explainability}} & \multicolumn{8}{c}{\textbf{Text Quality}} \\ \cline{2-13}
		& FMR$\uparrow$ & FCR$\uparrow$ & DIV$\downarrow$ & USR$\uparrow$ & B1$\uparrow$ & B4$\uparrow$ & R1-P$\uparrow$ & R1-R$\uparrow$ & R1-F$\uparrow$ & R2-P$\uparrow$ & R2-R$\uparrow$ & R2-F$\uparrow$ \\ \hline \hline
				
		& \multicolumn{12}{|c}{Yelp} \\ \hline
		ACMLM & \textbf{0.05} & \textbf{0.31} & \textbf{0.95} & \textbf{0.95} & 7.01 & 0.24 & 7.89 & 7.54 & 6.82 & 0.44 & 0.48 & 0.39 \\
		\textbf{PEPLER-D} &\textbf{0.05}&0.24&1.53&0.13&\textbf{9.17}**&\textbf{0.40}**&\textbf{15.67}**&\textbf{10.47}**&\textbf{11.73}**&\textbf{1.09}**&\textbf{0.78}**&\textbf{0.83}** \\ \hline
		NRT & 0.06 & 0.12 & 1.67 & 0.20 & 10.92 & 0.60 & 16.73 & 11.91 & 12.89 & 1.63 & 1.21 & 1.26 \\
		Att2Seq & 0.05 & 0.05 & 2.25 & 0.05 & 10.25 & 0.54 & 17.13 & 11.44 & 12.72 & 1.49 & 1.13 & 1.16 \\
		PETER & \textbf{0.08} & 0.15 & 1.62 & 0.15 & 10.74 & 0.63 & 16.18 & 11.90 & 12.63 & 1.60 & 1.32 & 1.28 \\
		\textbf{PEPLER} & \textbf{0.08} & \textbf{0.30}** & \textbf{1.52} & \textbf{0.35}** & 11.23 & 0.73 & 17.51 & 12.55 & 13.53 & \textbf{1.86}* & 1.42 & 1.46 \\
		\textbf{PEPLER (MLP)} &\textbf{0.08}&0.24&1.58&0.25&10.95&0.68&\textbf{17.52}&12.31&13.36&1.83&1.34&1.40 \\
		\textbf{PEPLER (MF)} &\textbf{0.08}&0.27&1.66&0.30&\textbf{11.70}&\textbf{0.75}**&\textbf{17.52}&\textbf{12.85}**&\textbf{13.72}**&\textbf{1.86}*&\textbf{1.45}*&\textbf{1.48}** \\ \hline \hline
		
		& \multicolumn{12}{|c}{Amazon} \\ \hline
		ACMLM & \textbf{0.10} & \textbf{0.31} & 2.07 & \textbf{0.96} & 9.52 & 0.22 & 11.65 & 10.39 & 9.69 & 0.71 & 0.81 & 0.64 \\
		\textbf{PEPLER-D} &0.08&0.19&\textbf{1.85}*&0.15&\textbf{10.94}**&\textbf{0.49}**&\textbf{16.31}**&\textbf{11.80}**&\textbf{12.80}**&\textbf{1.43}**&\textbf{1.13}**&\textbf{1.16}** \\ \hline
		NRT & 0.10 & 0.04 & 2.71 & 0.09 & 12.06 & 0.69 & 17.17 & 13.15 & 13.83 & 1.94 & 1.68 & 1.64 \\
		Att2Seq & 0.09 & 0.04 & 2.64 & 0.05 & 12.07 & 0.73 & 18.35 & 12.86 & 14.14 & 2.01 & 1.56 & 1.61 \\
		PETER & 0.09 & 0.09 & 2.16 & 0.20 & 11.75 & 0.89 & 16.51 & 13.10 & 13.55 & 1.96 & 1.76 & 1.68 \\
		\textbf{PEPLER} & 0.11 & 0.27 & \textbf{2.06} & 0.38 & 13.19 & 1.05 & \textbf{18.51} & 14.16 & 14.87 & \textbf{2.36}* & 1.88 & 1.91 \\
		\textbf{PEPLER (MLP)} &0.11&\textbf{0.34}**&2.10&\textbf{0.48}**&\textbf{13.59}**&\textbf{1.08}**&17.94&\textbf{14.50}*&14.82&2.29&\textbf{1.96}*&\textbf{1.93}** \\
		\textbf{PEPLER (MF)} &\textbf{0.12}*&0.24&2.18&0.35&13.46&1.02&18.30&14.37&\textbf{14.92}*&2.29&1.92&1.90 \\ \hline \hline
		
		& \multicolumn{12}{|c}{TripAdvisor} \\ \hline
		ACMLM & \textbf{0.07} & \textbf{0.41} & \textbf{0.78} & \textbf{0.94} & 3.45 & 0.02 & 4.86 & 3.82 & 3.72 & 0.18 & 0.20 & 0.16 \\
		\textbf{PEPLER-D} &0.05&0.22&2.69&0.08&\textbf{14.61}**&\textbf{0.87}**&\textbf{18.07}**&\textbf{14.83}**&\textbf{15.32}**&\textbf{1.76}**&\textbf{1.66}**&\textbf{1.58}** \\ \hline
		NRT & 0.05 & 0.02 & 6.07 & 0.00 & 13.76 & 0.80 & 19.01 & 14.57 & 15.58 & 2.10 & 1.59 & 1.68 \\		
		Att2Seq & 0.06 & 0.05 & 4.74 & 0.02 & 15.20 & 0.96 & 18.74 & \textbf{16.42} & 16.38 & 2.42 & \textbf{2.32} & 2.19 \\
		PETER & \textbf{0.07} & 0.09 & 3.62 & 0.05 & 15.13 & 1.00 & 18.30 & 16.15 & 16.00 & 2.24 & 2.23 & 2.06 \\
		\textbf{PEPLER} & \textbf{0.07} & \textbf{0.21}** & \textbf{2.71}** & \textbf{0.24}** & 15.49 & 1.09 & 19.48 & 15.67 & 16.24 & 2.48 & 2.21 & 2.16 \\
		\textbf{PEPLER (MLP)} &\textbf{0.07}&0.10&3.33&0.08&15.70&1.04&18.87&16.21&16.24&2.35&2.26&2.12 \\
		\textbf{PEPLER (MF)} &\textbf{0.07}&\textbf{0.21}**&2.89&0.21&\textbf{16.02}&\textbf{1.15}*&\textbf{19.82}&16.31&\textbf{16.69}&\textbf{2.53}&\textbf{2.32}&\textbf{2.22} \\ \hline \hline
	\end{tabular}
\end{table}

\section{Results and Analysis} \label{sec:results}

In this section, we first quantitatively compare the performance of different explanation methods with automatic metrics.
We then further study the effect of our proposed two training strategies.
Next, we qualitatively examine two explanation samples as generated by all the methods.
After that, we visualize our method's attention weights to demonstrate that IDs can indeed be fused into the pre-trained model.
At last, we study the effect of model size on explanation generation performance.

\subsection{Quantitative Analysis on Explanations} \label{sec:exp}

The performance comparison between different explanation generation methods is shown in Table \ref{table:explanation}.
These methods are divided into two groups.
We first examine those that map IDs onto item features, i.e., ACMLM and PEPLER-D.
Our PEPLER-D consistently and significantly outperforms ACMLM on the three datasets in terms of text quality measured by BLEU and ROUGE.
This demonstrates its effectiveness in generating high-quality sentences that are semantically close to the ground-truth text.
Also, we notice that the performance gap between our PEPLER-D and ACMLM (a fine-tuned BERT) is extremely large, because the latter's generation is achieved by predicting masked tokens, which is quite different from conventional auto-regressive generation.
This may explain why ACMLM produces diverse sentences (high USR) and features (low DIV).
However, they could be less useful to real users and might even hurt user experience, since their text quality cannot be guaranteed (see the generated examples in Table \ref{table:case}).

Next, we analyze the results of models that directly leverage user and item IDs for explanation generation, i.e., NRT, Att2Seq, PETER and PEPLER.
As we can see, the text quality of these methods are largely improved compared with those that convert IDs into item features (i.e., ACMLM and PEPLER-D), because the conversion process may lose certain information of IDs, e.g., identification.
Among the four ID-based methods, NRT and Att2Seq generally achieve the same performance on all metrics, but neither of them are as comparable as PETER and PEPLER. 
Because NRT and Att2Seq are based on recurrent neural networks (i.e., GRU or LSTM), they may suffer from the notorious long-term dependency problem, and thus their sequence modeling capability could be impaired.
As a comparison, PETER and PEPLER do not have such an issue, since in Transformer future tokens at any time step are given access to all the past tokens.
Moreover, given the fact that PETER is a small unpretrained Transformer, it does not outperform PEPLER that is pre-trained on large textual corpora and hence possesses rich linguistic knowledge.
In the meantime, it proves the rationale of our continuous prompt learning approach that could effectively make use of such knowledge for generating better explanations.

We then make a comparison for our proposed two training strategies.
The default PEPLER employs sequential tuning, while the other two variants utilize recommendation as regularization with MLP and MF, respectively, and therefore are denoted as PEPLER (MLP) and PEPLER (MF).
Compared with PEPLER, PEPLER (MF) greatly improves the text quality most of the time.
In the meantime, PEPLER (MLP) maintains comparable text quality to PEPLER, but often cannot keep up explainability, e.g., the decrease on FCR and USR.
This can be explained by the difference between MF and MLP in terms of additional parameters for recommendation task.
For MF, the prediction is simply made by the dot product between user and item embeddings, in which case no additional parameters are involved.
In contrast, MLP must go through a stack of hidden layers that consist of many parameters, which might help to predict ratings but adversely affect the learning of the explanation task.
Since the recommendation task requires extra rating data for training, which may not always be available in other natural language generation tasks (e.g., dialogue systems), we set sequential tuning as the default training strategy for PEPLER.
Depending on the specific application, one may consider PEPLER (MF).

From the experimental results, we also observe two special cases on the TripAdvisor dataset, where Att2Seq obtains the largest ROUGE scores.
The reasons are as follows.
First, we fixed its generation issue (see the discussion in Section \ref{sec:methods}), which makes it a competitive baseline.
Second, the dataset is quite small and thus the training samples are limited, so our large model may underfit.
This is not a problem in real-world applications where there are abundant training samples (e.g., in e-commerce), since our model already outperformed state-of-the-art baselines on the largest dataset Yelp, which contains approximately 1.3 million samples.

\begin{figure}
	\centering
	\subfigure{\includegraphics[scale=0.45]{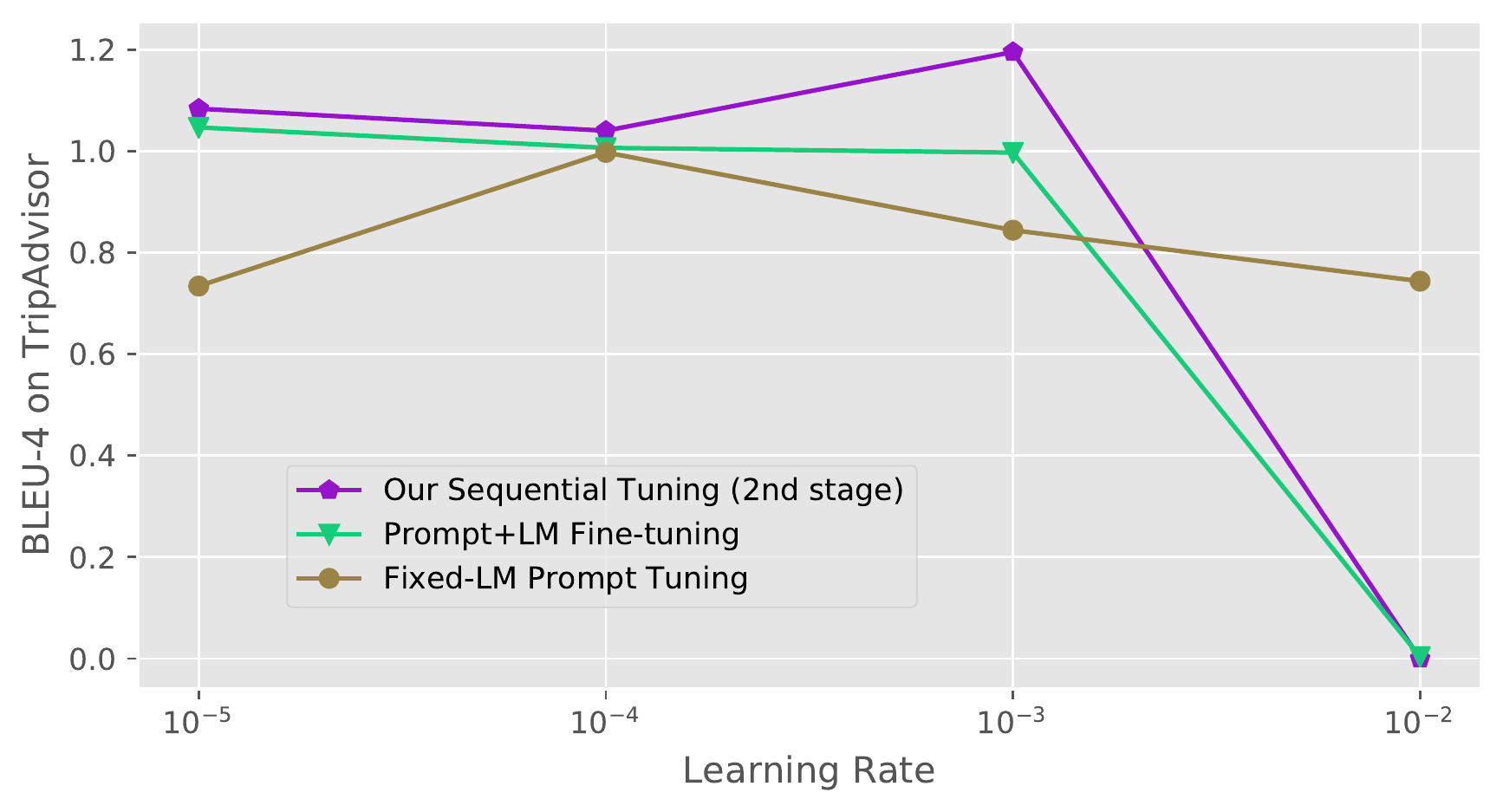}}
	\subfigure{\includegraphics[scale=0.45]{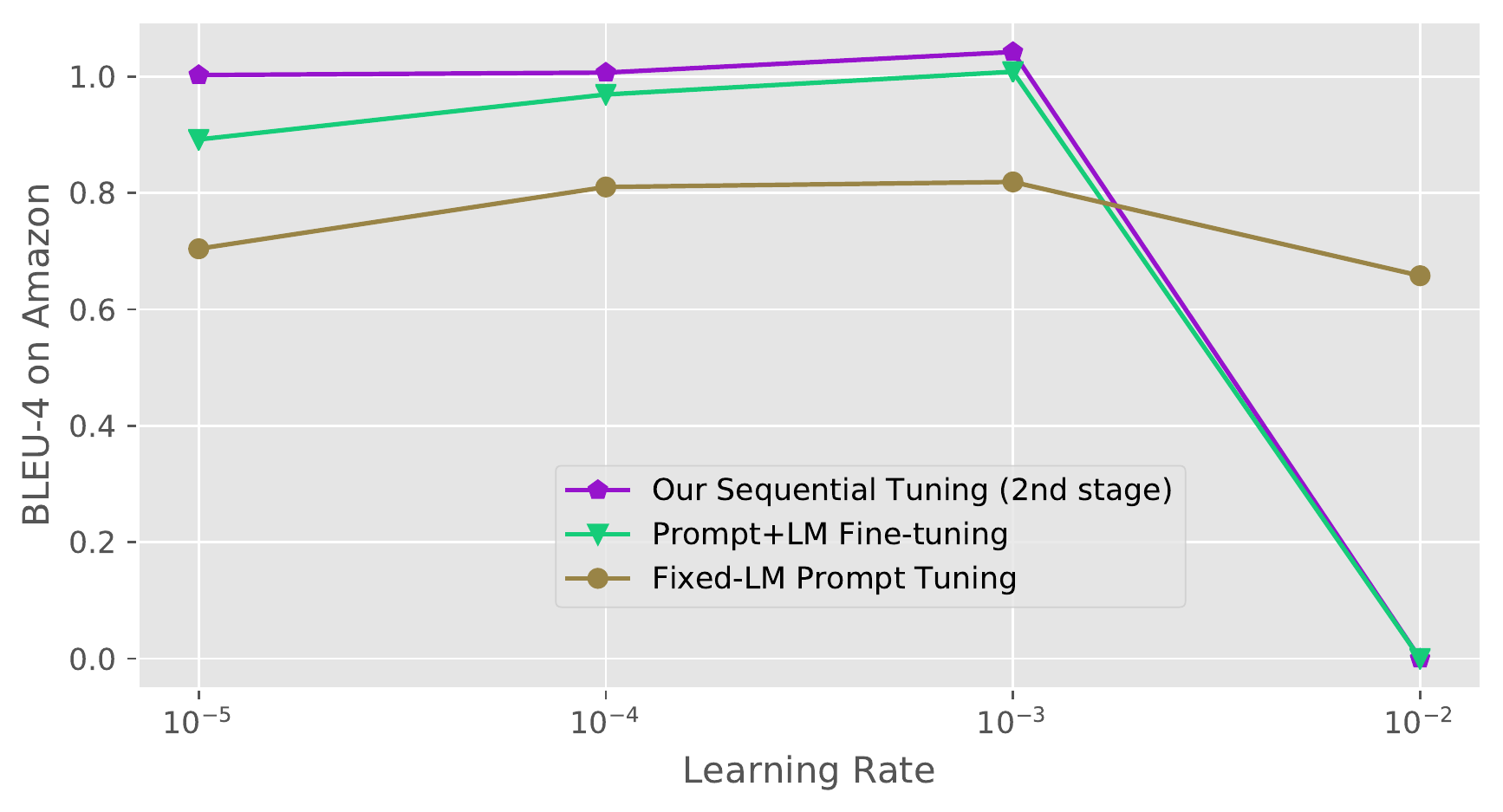}}
	\subfigure{\includegraphics[scale=0.45]{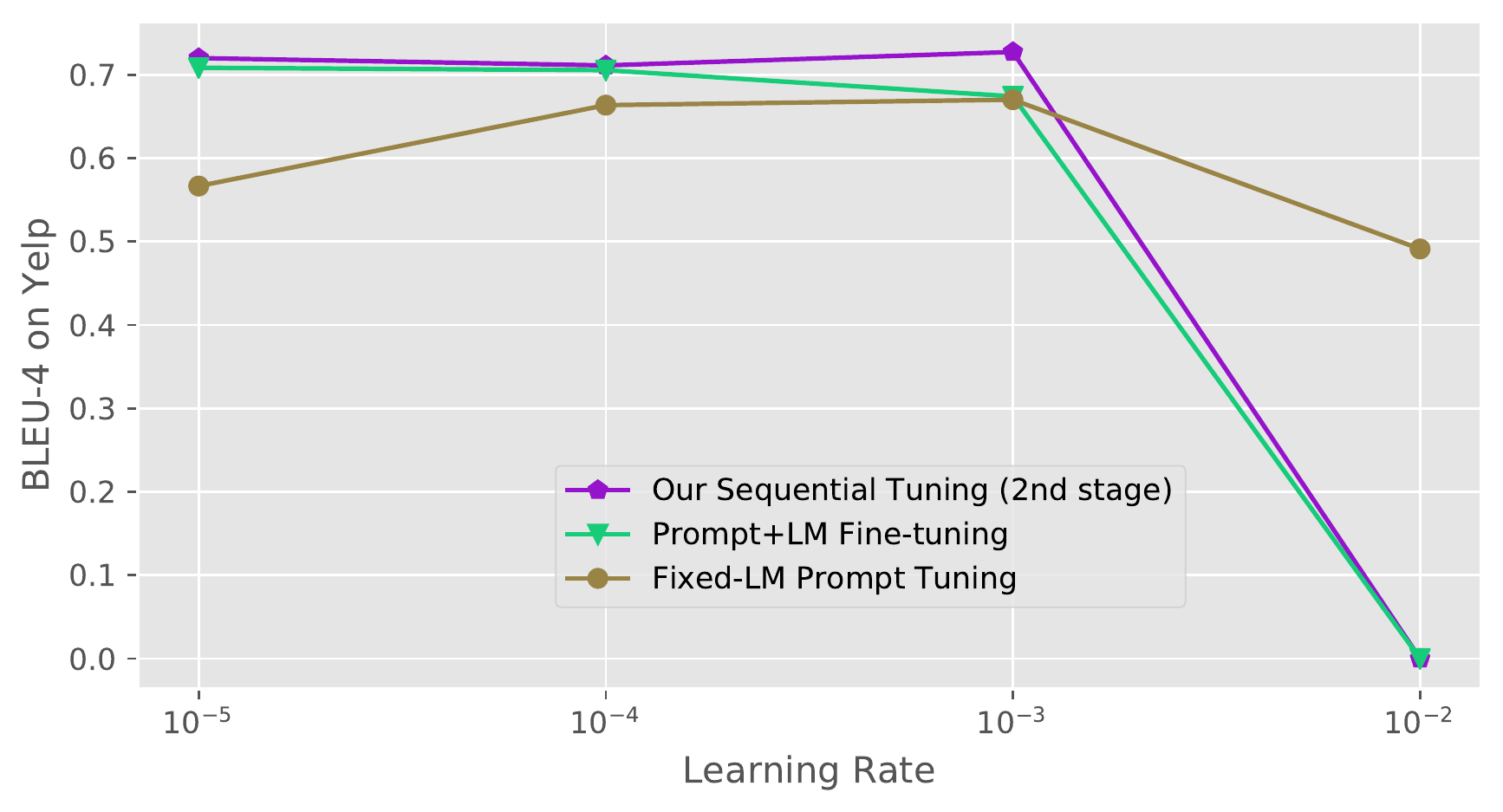}}
	\caption{A comparison of three tuning strategies for continuous prompt learning in terms of BLEU-4 with varying learning rates on three datasets.}
	\label{fig:lr}
\end{figure}

\subsection{Effect of Sequential Tuning} \label{sec:sequential}

To validate the superiority of our proposed Sequential Tuning strategy, we compare it with its two composite training strategies: Fixed-LM Prompt Tuning and Prompt+LM Fine-tuning \cite{CSUR22-Survey}.
The results of Sequential Tuning (utilized in the default PEPLER) on the three datasets are presented in Table \ref{table:explanation}.
Given the consistent performance across different metrics, in Fig. \ref{fig:lr} we only show BLEU-4 with varied learning rates on three datasets.

As it can be seen, the highest BLEU-4 score is achieved by our Sequential Tuning strategy (purple), when the learning rate is set to $10^{-3}$.
This manifests its advantage in bridging the gap between the randomly initialized continuous prompts and the pre-trained language model.
In particular, the pattern of our Sequential Tuning and that of Prompt+LM Fine-tuning (green) is quite similar, because they both tune all the model parameters, including both prompts and the pre-trained model.
Obviously, the curve of our Sequential Tuning is on the top of that of Prompt+LM Fine-tuning.
The difference is that the former's prompts are already trained, which could help to reduce the gap between prompts and the pre-trained model.
This supports the rationale of our two-staged Sequential Tuning strategy.
Moreover, when the learning rate is large (i.e., $10^{-2}$), the performance of both strategies goes down dramatically, nearly reaching 0, because large learning rates lead to significant changes of parameters in the pre-trained model.
Hence, smaller learning rates are more appreciated to fine-tuning.
In contrast, the performance of Fixed-LM Prompt Tuning (brown) is relatively stable, regardless of the changing learning rates.
However, it does not outperform the other two strategies, because the model is frozen and only prompts can be tuned, and therefore could not be well adjusted to the target explanation task.

\subsection{Effect of Recommendation as Regularization} \label{sec:regularization}

In this work, we propose two training strategies to bridge continuous prompts and the pre-trained model, including sequential tuning and recommendation as regularization.
We analyze the latter in more details, because the former is already presented in the previous subsection.

\begin{figure}
	\centering
	\subfigure[BLEU-4 with varying $\lambda$]{\includegraphics[scale=0.4]{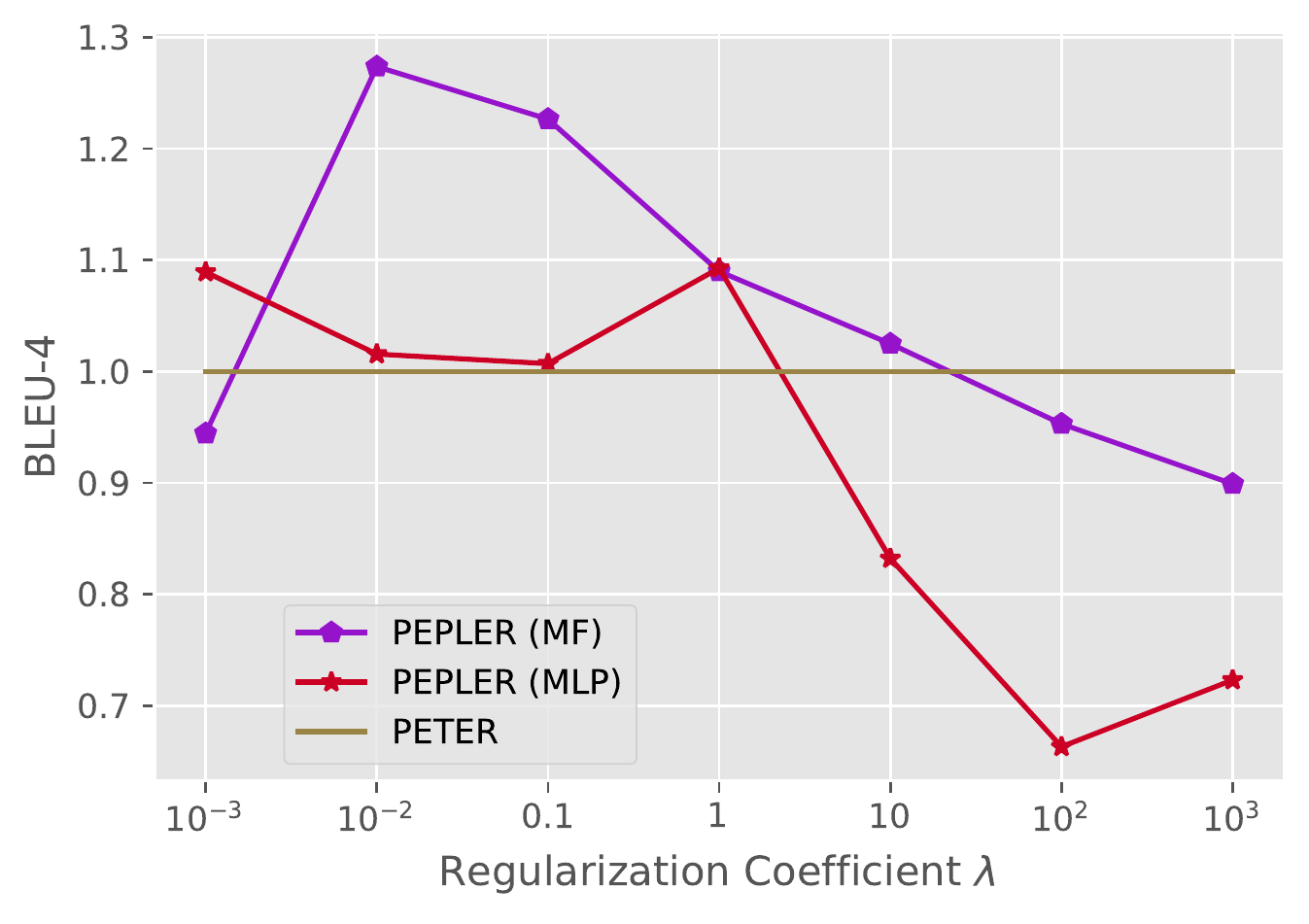}}
	\subfigure[RMSE with varying $\lambda$]{\includegraphics[scale=0.4]{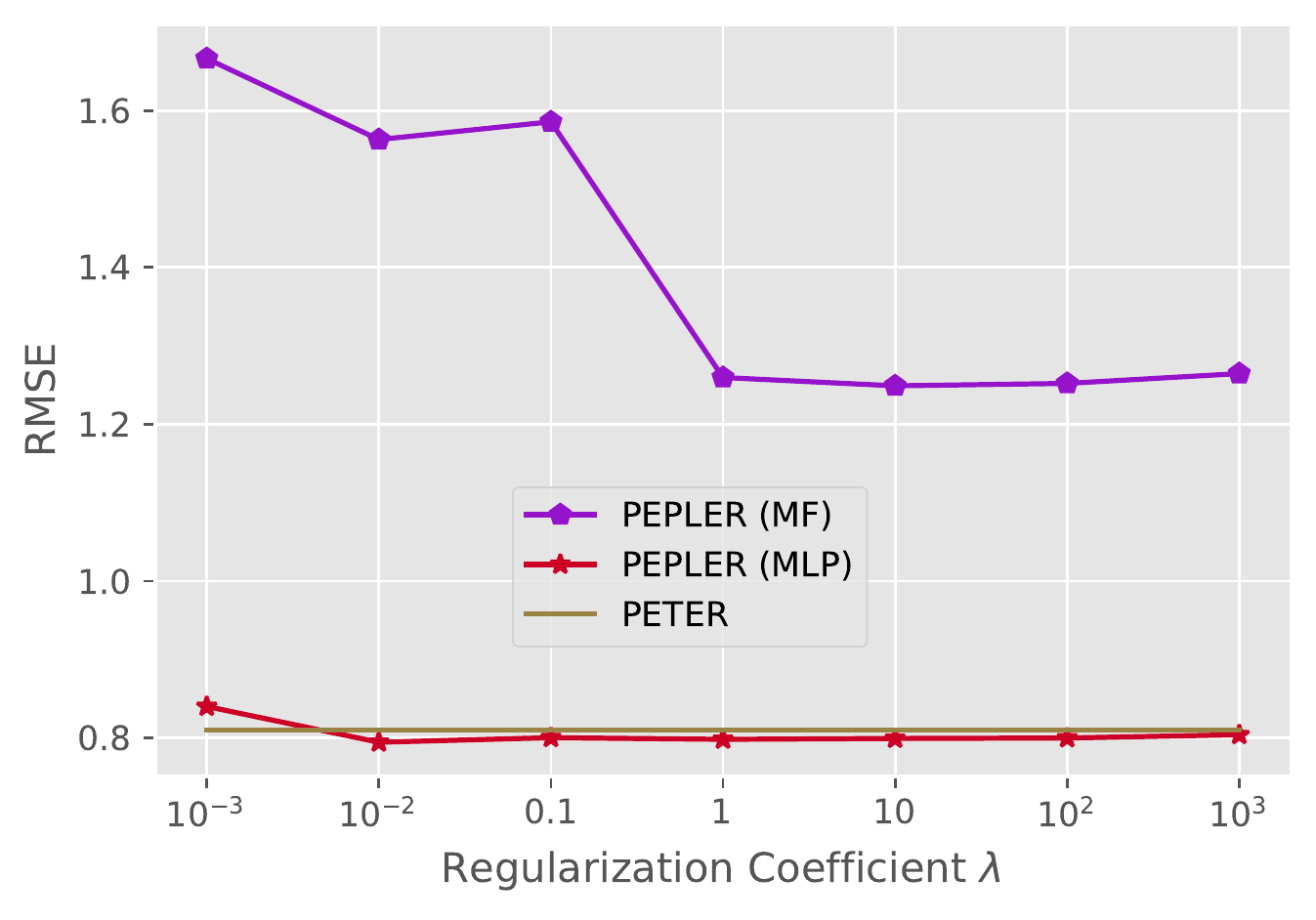}}
	\subfigure[USR with varying $\lambda$]{\includegraphics[scale=0.4]{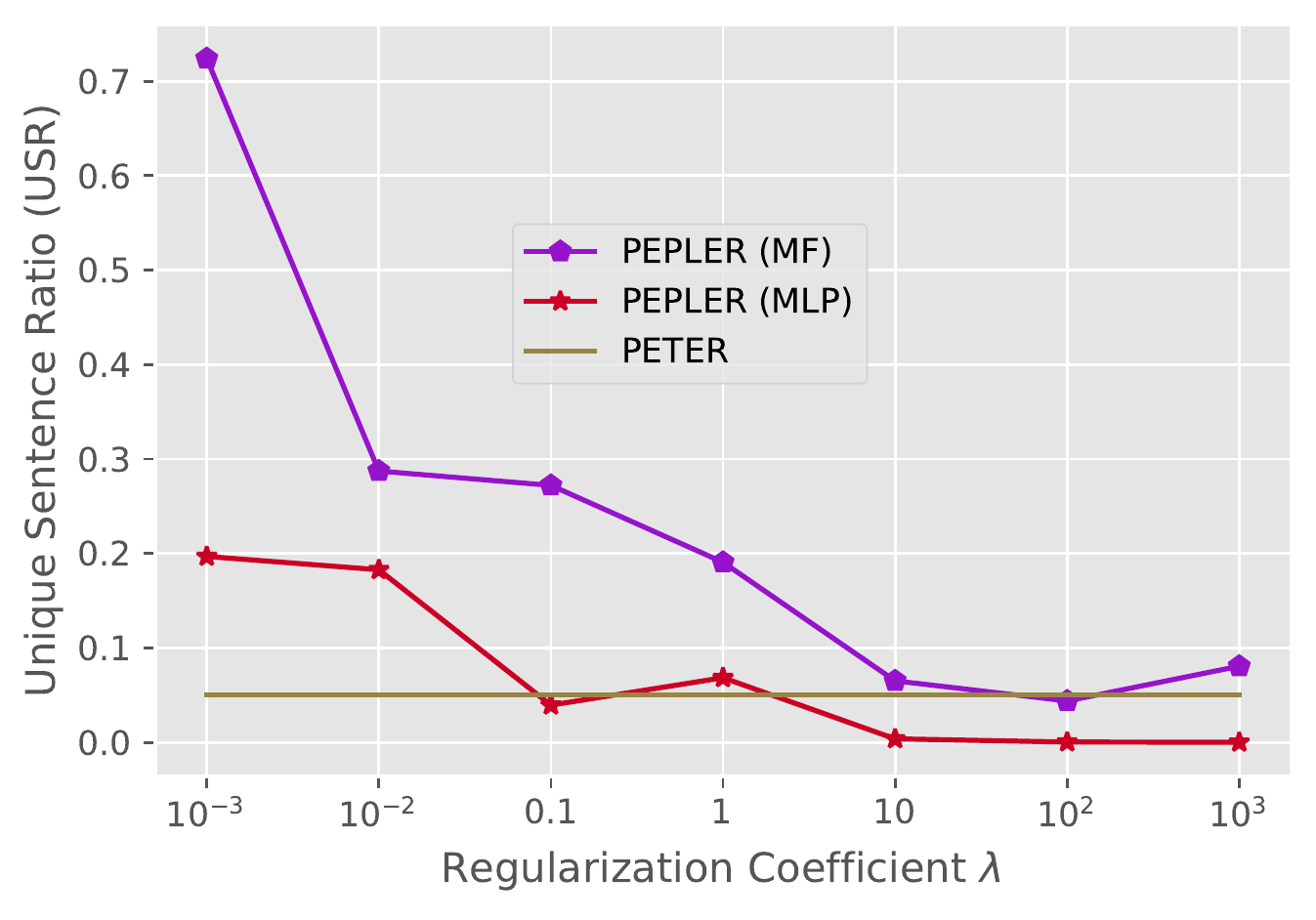}}
	\subfigure[FCR with varying $\lambda$]{\includegraphics[scale=0.4]{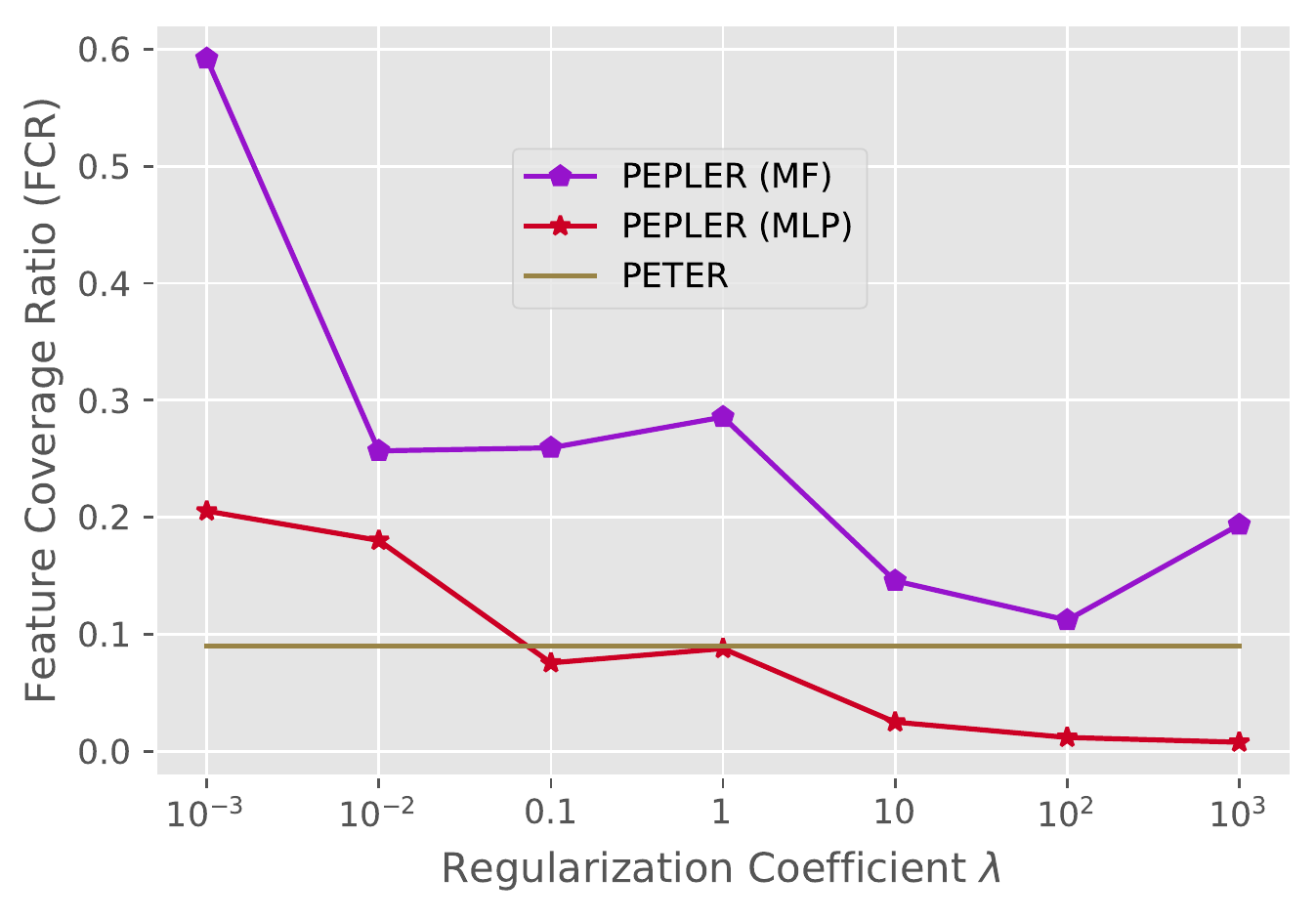}}	
	\caption{The effect of regularization coefficient $\lambda$ on the recommendation task with MF or MLP for PEPLER on the TripAdvisor dataset. For better comparison, the results of PETER are shown.}
	\label{fig:reg}
\end{figure}

In Fig. \ref{fig:reg}, we investigate how PEPLER (MF) and PEPLER (MLP) react to varying $\lambda$, the regularization coefficient on the recommendation task.
For better comparison, PETER is included since it is the previous state-of-the-art, and can also perform recommendation.
The accuracy of this task is measured by root mean square error (RMSE), and a lower score indicates a better performance.
By comparing the first two sub-figures, we can clearly see that there is a trade-off between explanation text quality (evaluated by BLEU-4) and recommendation accuracy (measured by RMSE) for PEPLER (MF).
For example, when $\lambda = 10^{-2}$, its explanation performance reaches an optimal, but its recommendation performance is greatly deteriorated.
It actually supports our design of this training strategy that leverages the recommendation task to help the learning of explanation generation.
As a comparison, PEPLER (MLP) is not so sensitive to varying $\lambda$.
We also notice that there is a huge gap between PEPLER (MF) and PEPLER (MLP) in terms of recommendation accuracy.
Owing to the linearity of MF, its representation ability could be largely limited \cite{WWW17-NeuMF}, and thus could not accurately estimate the ratings.
But because of the simple dot product operation, the relation between users and items encoded in ratings could in turn be easily propagated to better learn the explanation task, i.e., higher BLEU-4 for PEPLER (MF).
Since the purpose of PEPLER (MF) is not to make recommendations, when deploying it for real-world applications, one can use the predictions from another effective recommendation model, e.g., neural matrix factorization \cite{WWW17-NeuMF}.

The last two sub-figures show a decline of explainability as measured by Unique Sentence Ratio (USR) and Feature Coverage Ratio (FCR) for both PEPLER (MF) and PEPLER (MLP), with the increase of $\lambda$.
It suggests that a smaller $\lambda$ could lead to larger USR and FCR.
However, this pattern does not match that of text quality as measured by BLEU-4.
When text quality cannot be guaranteed, the explanations could be unreadable to users and thus may affect their experience.
In such cases, large explainability scores would be pointless.
Therefore, we give priority to text quality when tuning $\lambda$ for both PEPLER (MF) and PEPLER (MLP).

\subsection{Qualitative Case Study on Explanations}

\begin{table}
	\caption{Explanations on two different cases as generated by different methods on the TripAdvisor dataset. Special tokens used to perform generation (i.e., $<$\textit{bos}$>$ and $<$\textit{eos}$>$) are removed for the ease of readability. The boldfaced words in the ground-truth are the key features. Matched features in the generated explanations are also boldfaced.}
	\label{table:case}
	\centering
	\begin{tabular}{l|p{11cm}}
		\hline \hline
		Ground-truth & the swimming \textbf{pool} is fantastic \\
		\hline
		ACMLM & swimming \textbf{pool} swimming \textbf{pools} \textbf{pool} strip beach area \\		
		NRT & the hotel is located in a great location \\
		\multirow{2}{*}{Att2Seq} & the hotel is located in the heart of the city and the main shopping area is also within walking distance \\
		PETER & the hotel is located in the heart of the city and the harbour \\
		\textbf{PEPLER-D} & the room was very nice and the bed was very comfortable \\
		\textbf{PEPLER} & the \textbf{pool} is amazing and the \textbf{pool} is very relaxing \\
		\hline \hline
		Ground-truth & this is one of the finest \textbf{hotels} in all of Europe \\
		\hline
		\multirow{2}{*}{ACMLM} & swimming pool area pool ja \#\#cu \#\#zzi pool city area gym building pool area spa gym pool area \\		
		NRT & the \textbf{hotel} is located in a great location \\
		\multirow{2}{*}{Att2Seq} & the \textbf{hotel} is located in the heart of the city and the main shopping area is also within walking distance \\
		PETER & the \textbf{hotel} is in a great location \\
		\textbf{PEPLER-D} & the \textbf{hotel} is a short walk from the old town \\
		\textbf{PEPLER} & the \textbf{hotel} is located in the heart of the city and is very well maintained \\
		\hline \hline
	\end{tabular}
\end{table}

In Table \ref{table:case}, we present two examples generated by all the methods for hotel recommendations on the TripAdvisor dataset.
In the first case, the ground-truth explanation gives a positive comment about the hotel's swimming ``pool''.
Only two methods, i.e., ACMLM and our PEPLER, successfully capture this key feature.
However, ACMLM's explanation is not even readable, because it is just a bunch of unordered random words.
These meaningless explanations are not very likely to be useful to real users.
As a comparison, the explanations generated by the other approaches are all readable and fluent.
This actually echoes their performances on BLEU and ROUGE, which emphasize more text quality and readability.
But BLEU and ROUGE are not perfect, because they fail to detect the problem of identical explanations (see the same sentences generated by NRT or Att2Seq for two different cases).
This is why we also adopt some explainability metrics \cite{CIKM20-NETE} that particularly care about item features and sentence diversity.
Moreover, Att2Seq tends to generate long explanations, which may explain why it obtains good performance regarding ROUGE on the TripAdvisor dataset (see Table \ref{table:explanation}), because ROUGE is a recall-oriented metric and favors long sentences.
The explanations generated by the other three approaches, i.e., PETER, PEPLER-D and PEPLER, are quite good, because they all adopt the Transformer model, which has strong language modeling capability.
Despite of that, the explanations from our PEPLER are semantically closer to the ground-truth.
Taking the second case as an example, the ground-truth explanation evaluates the overall quality of the hotel (``one of the finest hotels''), but PETER and PEPLER-D respectively talks about location (``great location'') and distance (``short walk''), while our PEPLER comments about not only the hotel's location (``located in the heart of city'') but also its quality (``well maintained'').
We attribute this to the effectiveness of our proposed continuous prompt learning and the sequential tuning strategy.
Moreover, we see that the expression of PEPLER's explanations is quite rich, which could be brought by the linguistic knowledge contained in the pre-trained model, as it is already trained on large text corpora.

\subsection{Attention Visualization}

\begin{figure}
	\centering
	\subfigure[Before training, the model could not make use of user and item IDs for explanation generation.]{\includegraphics[scale=0.42]{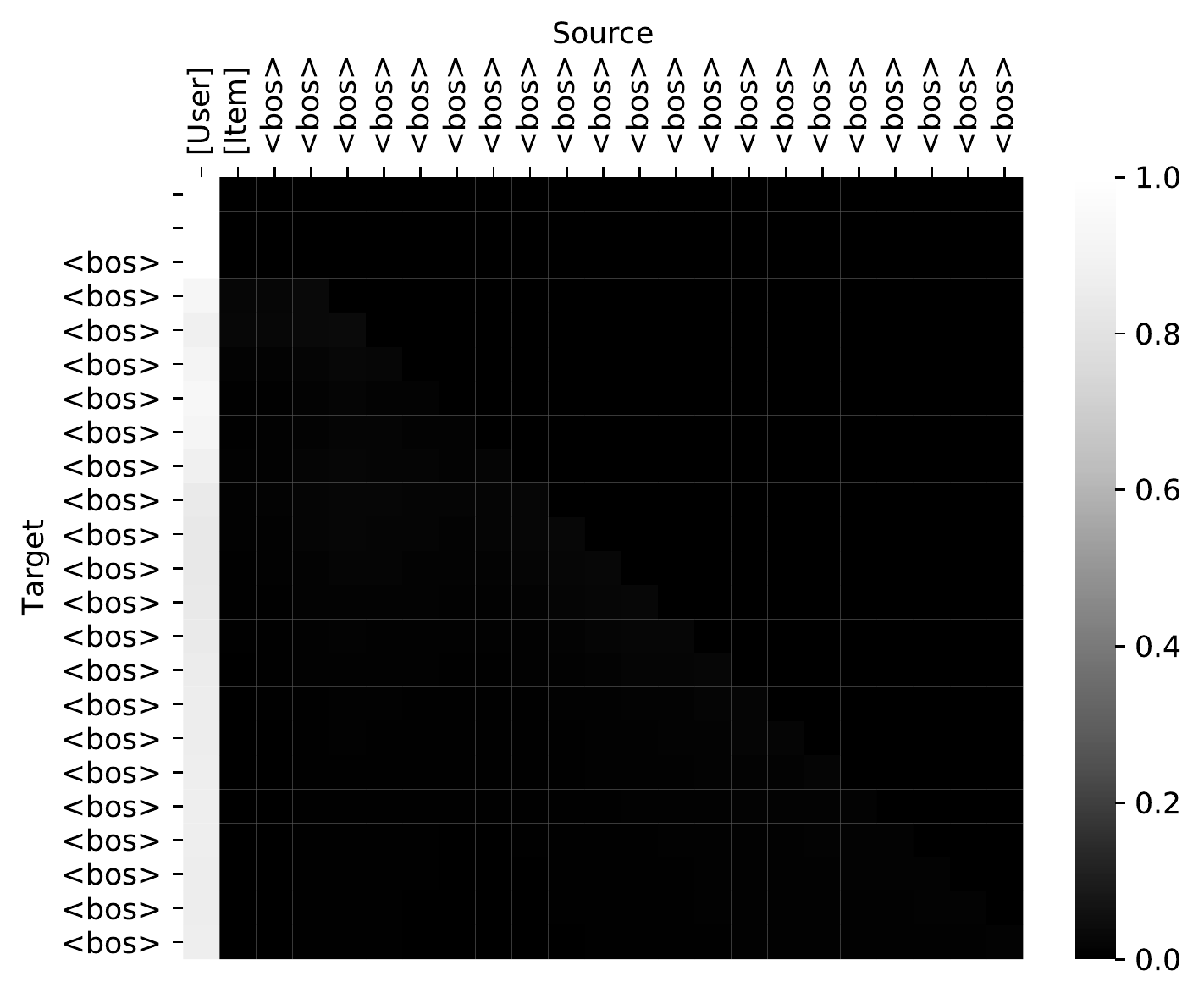}}
	\hspace{10mm}
	\subfigure[After training, the IDs can be effectively utilized by our model (see the first two columns).]{\includegraphics[scale=0.42]{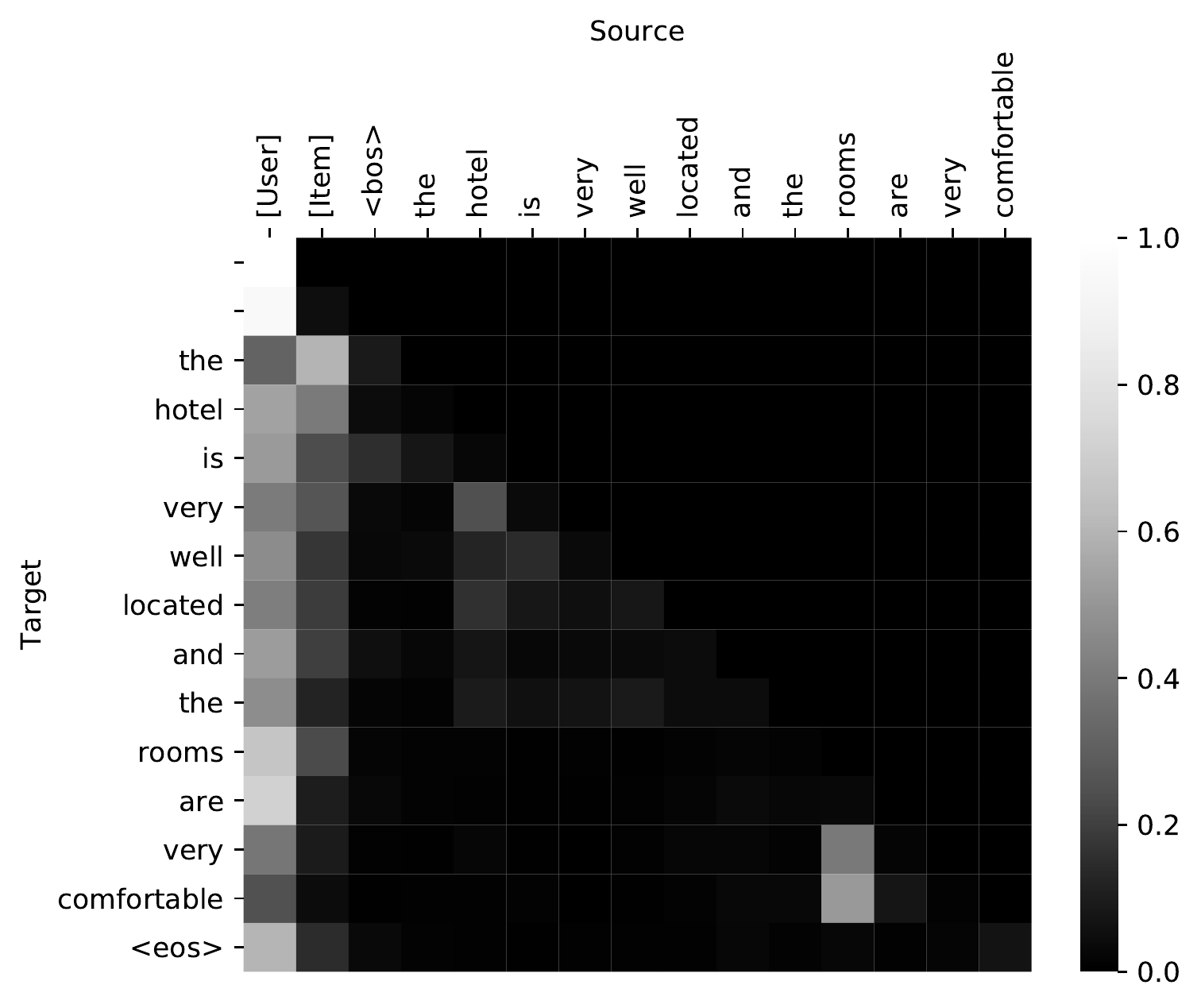}}
	\caption{Visualization of our PEPLER model's last attention layer, before and after training. The larger the attention weights, the lighter the cells.}
	\label{fig:attention}
\end{figure}

In our continuous prompt learning approach, we directly incorporate user and item IDs into the pre-trained model for natural language explanation generation for recommendations.
To see whether the IDs are really fused into the model, we visualize its last attention layer before and after training in Fig. \ref{fig:attention}.
In both sub-figures, the larger an attention weight, the lighter the corresponding cell.
Before training, the ID representations are randomly initialized, but the model is already trained on large textual corpora.
This semantic gap makes the pre-trained model difficult to perform natural language generation based on IDs.
From Fig. \ref{fig:attention} (a), we can see that the model cannot utilize both user and item IDs before training, resulting in an unreadable sequence of multiple $<$\textit{bos}$>$.
But after training, the model is able to make use of the IDs and thus can generate a fluent and readable explanation, e.g., ``the hotel is very well located and the rooms are very comfortable''.
It confirms that the IDs can indeed be well fused into the model.
We attribute this to the effectiveness of our proposed sequential tuning approach.

\subsection{Effect of Model Size}

\begin{figure}
	\centering
	\includegraphics[scale=0.7]{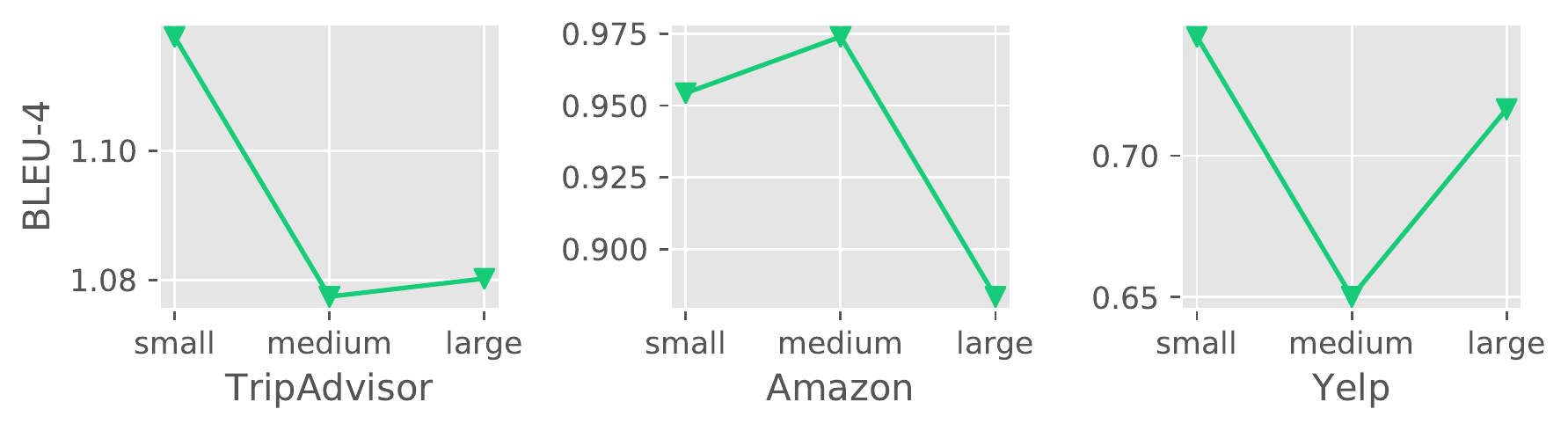}
	\caption{The effect of model size on text quality in terms of BLEU-4 on three datasets.}
	\label{fig:size}
\end{figure}

The pre-trained GPT-2 model \cite{19-GPT2} has four varying sizes, including Small, Medium, Large and XL.
This work is based on the default 12-layered small model, while the others have 24, 36, and 48 layers, respectively.
Here, we investigate whether larger models with more attention layers could lead to better explanation generation performance.
In Fig. \ref{fig:size}, we present their text quality as measured by BLEU-4 on the three datasets, where the XL model is omitted because it is too large and ran out of memory in our every experimental trial.
From the three sub-figures, we do not observe an increasing trend with the increase of model size, and therefore cannot certify that a larger model always leads to a better performance.
We conjecture that large models might suffer from data-hungry problem and therefore may need more data to perform well.
Nevertheless, the small model consistently reaches a reasonably good performance on three datasets, while it has less model parameters and thus takes less time to fine-tune.
It actually supports our choice of the default model.

\section{Conclusion} \label{sec:conclude}

In this work, we propose two prompt learning approaches to exploit the rich knowledge contained in pre-trained language models for recommendation explanation generation.
To bridge the gap between continuous prompts and pre-trained models, we come up with two effective learning strategies.
Extensive experiments demonstrate the effectiveness of our approaches in generating high-quality explanations as measured by text quality and explainability metrics.

As future works, we are immensely interested in whether the generated explanations possess bias or stereotype against certain groups of users and how to mitigate them, as reported in recent studies \cite{EMNLP19-Bias, ICML21-Bias}, pre-trained models may exhibit societal bias towards different demographics.
Moreover, since the biased generation was triggered by discrete prompts \cite{EMNLP19-Bias}, we wonder whether it is possible to design some other discrete prompts that can help us diagnose the behavior of pre-trained models, which would certainly increase their interpretability.
Besides explanation generation for recommender systems, we also plan to adopt our approaches to other applications of personalized natural language generation, such as personalized question answering systems and personalized conversational agents.
Moreover, it would also be interesting to incorporate item images into pre-trained models to generate visual explanations for recommendations, since ``a picture is worth a thousand words''.
Another meaningful extension is to adapt pre-trained models to cross-lingual explanation generation, since international platforms, e.g., Amazon, may serve users who speak different languages.

\begin{acks}
This work was supported by Hong Kong RGC GRF project (RGC/HKBU12201620), Hong Kong Baptist University IG-FNRA project (RC-FNRA-IG/21-22/SCI/01), and partially supported by NSF IIS-1910154, 2007907, and 2046457. 
Any opinions, findings, conclusions or recommendations expressed in this material are those of the authors and do not necessarily reflect those of the sponsors.
\end{acks}

\bibliographystyle{ACM-Reference-Format}
\bibliography{bibliography}

\appendix

\end{document}